\newif\iflabor
\labortrue
\makeatletter

\documentclass[peerreview,12pt]{IEEEtran}

\usepackage{mathtools,amssymb,amsthm}
    \allowdisplaybreaks
    \def\FF{\mathbf F}
    \def\PP{\mathbf P}
    \def\RR{\mathbf R}
    \def\II{\mathcal I}
    \def\leq{\leqslant}
    \def\geq{\geqslant}
    \def\loll{[^1_1{}^0_1]}
    \def\lowl{[^1_\omega{}^0_1]}
    \def\oooo{[^0_0{}^0_0]}
    \def\olll{[^0_1{}^1_1]}
    \def\lllo{[^1_1{}^1_0]}
    \def\lool{[^1_0{}^0_1]}
    \DeclareMathOperator\tr{tr}
    \DeclareMathOperator\BEC{BEC}
    \DeclareMathOperator\TEC{TEC}
    \DeclareMathOperator\Dual{Dual}
    \catcode`0=13
    \gdef\installlightgrayzero{%
        \catcode`0=13 \def0{{\color{lightgray}\char48}}
    }
    \catcode`0=12

\usepackage{pgfmath}
    \def\centerbox{\raisebox{\fontdimen22\textfont2}}
    \def\baselinestar{\raisebox{-0.228em}{$\star$}\rule[-3pt]{0pt}{6pt}}
    \def\scalestar#1{\scalebox{#1}{\baselinestar}}
    \def\rotatestar#1#2{\rotatebox[origin]{#1}{\scalestar{#2}}}
    \def\randomstar#1#2{\pgfmathsetmacro\a{rnd*360}\rotatestar{\a}{#2}}
    \def\centerstar#1#2{\centerbox{\randomstar{#1}{#2}}}
    \def\ostarprototype{\mathbin{\ooalign{%
        $\odot$\cr
        \hfil\color{white}$\bullet$\hfil\cr
        \hfil\centerstar{120}{0.8}\hfil\cr
    }}}
    \def\boxstarprototype{\mathbin{\ooalign{%
        $\phantom\odot$\cr
        \hfil\centerbox{\rule[-.3em]{.6em}{.6em}}\hfil\cr
        \hfil\centerbox{\color{.!0}\rule[-.265em]{.53em}{.53em}}\hfil\cr
        \hfil\centerstar{240}{0.8}\hfil\cr
    }}}
    \def\boxstar{\mathchoice
        {\boxstarprototype}
        {\boxstarprototype}
        {\scalebox{0.83}{$\boxstarprototype$}}
        {\scalebox{0.7}{$\boxstarprototype$}}
    }
    \def\ostar{\mathchoice
        {\ostarprototype}
        {\ostarprototype}
        {\scalebox{0.83}{$\ostarprototype$}}
        {\scalebox{0.7}{$\ostarprototype$}}
    }
    \def\parall{\ostar}
    \def\para{{\ostar}}
    \def\serial{\boxstar}
    \def\seri{{\boxstar}}

\usepackage{tikz,pgfplotstable}
    \definecolor{PMS3015}{HTML}{00629B}
    \definecolor{PMS1245}{HTML}{C69214}
    \tikzset{every picture/.style={line cap=round,line join=round}}
    \pgfplotsset{compat=1.17}
    \usetikzlibrary{decorations.text}

\usepackage[unicode, pdfusetitle]{hyperref}
    \hypersetup{
        pdfsubject={Information Theory (cs.IT)},
        pdfkeywords={polar code, scaling exponent, non-binary alphabet},
        colorlinks, allcolors=green!25!blue!66!black,
    }

\usepackage{cleveref}
    
        \crefname{section}{Section}{Sections}
        \crefname{appendix}{Appendix}{Appendices}
        \crefname{figure}{Figure}{Figure}
        \crefname{table}{Table}{Tables}
    
    \def\cthm#1#2#3{\newtheorem{#1}[allamsthm]{#2}\crefname{#1}{#2}{#3}}
        \cthm{lemma}{Lemma}{Lemmas}
        \cthm{theorem}{Theorem}{Theorems}
        \cthm{proposition}{Proposition}{Propositions}
        \cthm{corollary}{Corollary}{Corollaries}

\title{
                Accelerating Polarization via Alphabet Extension
}
\author{
                                  Iwan Duursma
                                    ~~and~~
                                  Ryan Gabrys
                                    ~~and~~
                              Venkatesan Guruswami
                                    \\and~~
                                 Ting-Chun Lin
                                    ~~and~~
                                  Hsin-Po Wang%
\thanks{
    This work was presented at 2022 International Conference on
    Randomization and Computation (RANDOM) \cite{TetraErase22c}.  Iwan
    Duursma is with University of Illinois Urbana-Champaign, IL, USA;
    email: duursma@illinois.edu; orcid: 0000-0002-2436-3944; he is
    supported by Simons Foundation grant 714010.  Ryan Gabrys is with
    University of California San Diego, CA, USA; email:
    rgabrys@ucsd.edu; orcid: 0000-0002-9197-3371; he is supported by NSF
    grant CCF-2107346.  Venkatesan Guruswami is with University of
    California, Berkeley, CA, USA; email: venkatg@berkeley.edu; orcid:
    0000-0001-7926-3396; he is supported by NSF grant CCF-2210823 and
    Simons Investigator Award.  Ting-Chun Lin is with University of
    California San Diego, CA, USA and Hon Hai (Foxconn) Research
    Institute, Taipei, Taiwan; email: til022@ucsd.edu; orcid:
    0000-0002-8994-4598.  Hsin-Po Wang was with University of California
    San Diego, CA, USA; he is now with University of California,
    Berkeley, CA, USA.  email: simple@berkeley.edu; orcid:
    0000-0003-2574-1510; he is supported by NSF grant CCF-1764104.
}}

\begin{document}

\maketitle

\begin{abstract}\def\GF(#1){\mathbb F_{#1}}%
    Polarization is an unprecedented coding technique in that it not
    only achieves channel capacity, but also does so at a faster speed
    of convergence than any other coding technique.  This speed is
    measured by the ``scaling exponent'' and its importance is
    three-fold.  Firstly, estimating the scaling exponent is challenging
    and demands a deeper understanding of the dynamics of communication
    channels.  Secondly, scaling exponents serve as a benchmark for
    different variants of polar codes that helps us select the proper
    variant for real-life applications.  Thirdly, the need to optimize
    for the scaling exponent sheds light on how to reinforce the design
    of polar codes.

    In this paper, we generalize the binary erasure channel (BEC), the
    simplest communication channel and the protagonist of many coding
    theory studies, to the ``tetrahedral erasure channel'' (TEC).  We
    then invoke Mori--Tanaka's $2 \times 2$ matrix over $\GF(4)$ to
    construct polar codes over TEC.  Our main contribution is showing
    that the dynamic of TECs converges to an almost--one-parameter
    family of channels, which then leads to an upper bound of $3.328$ on
    the scaling exponent.  This is the first non-binary matrix whose
    scaling exponent is upper-bounded.  It also polarizes BEC faster
    than all known binary matrices up to $23 \times 23$ in size.  Our
    result indicates that expanding the alphabet is a more effective and
    practical alternative to enlarging the matrix in order to achieve
    faster polarization.
\end{abstract}

\begin{IEEEkeywords}
    Polar code, scaling exponent, non-binary alphabet.
\end{IEEEkeywords}

\section{Introduction}

    A fundamental question at the center of the theory of communication
    is whether we can fully utilize a noisy channel to transmit
    information.  In modern terminology, can error correcting codes
    achieve channel capacity?  The answer is positive; in fact, multiple
    code constructions do so.  Among them, polar code is a special one
    as it achieves capacity faster than any other known code.

    Polar coding was invented by Arıkan around 2008 \cite{Ari09}.
    During that time, Arıkan was experimenting with channel combining
    and splitting.  By treating two independent binary channels as a
    single quaternary channel (combining) and tasking ourselves with
    guessing certain linear combinations of the inputs (splitting), he
    synthesized two channels, denoted by $W^\seri$ and $W^\para$, out of
    the original channel $W$.  Arıkan realized that, when combining and
    splitting is applied recursively, the channels undergo an intriguing
    dynamic that ultimately results in most synthetic channels being
    either almost noiseless or extremely noisy.  This is \emph{channel
    polarization}, the first ingredient underlying polar codes.

    The second ingredient of polar codes, also given by Arıkan in said
    seminal paper, is the relation between the dynamic of synthetic
    channels and the construction and performance of codes.  Arıkan's
    insight was that synthetic channels that become almost noiseless can
    be used to transmit information bits, and synthetic channels that
    become extremely noisy can be frozen to some fixed values.  The rate
    at which we communicate meaningful bits is then the proportion of
    synthetic channels that are almost noiseless.  So, whether we can
    achieve channel capacity becomes a problem of counting the number
    of good and bad synthetic channels.

    It then became apparent, perhaps even appealing, that one can study
    the dynamic of synthetic channels by means of stochastic processes.
    Take the \emph{binary erasure channel} (BEC) as an example.  Let $W$
    be $\BEC(\varepsilon)$, the BEC with erasure probability
    $\varepsilon$, where $0 < \varepsilon < 1$.  The synthetic channels
    $W^\seri$ and $W^\para$ are $\BEC(2\varepsilon - \varepsilon^2)$ and
    $\BEC(\varepsilon^2)$, respectively.  Accordingly, a stochastic
    process $\{H_n\}_n$ is defined by having $H_0 \coloneqq \varepsilon$
    and $H_{n+1} \coloneqq 2H_n - H_n^2$ or $H_n^2$ with equal
    probability.  It can be shown that if
    \[ \PP\{\, H_n \leq f(n) \,\} \geq 1 - H_0 - g(n), \]
    where $f$ and $g$ are functions in $n$, then there is a polar code
    with length $2^n$, miscommunication probability at most $2^n f(n)$,
    and gap to capacity at most $g(n)$.

    It was at this point that the study of polar codes thrived and
    branched.  On the \emph{error exponent} branch, $g$ is a constant
    and the asymptotics of $f$ is studied.  It was shown that $f(n)$ is
    roughly $\exp(-e^{\beta n})$, where $\beta > 0$ is a constant
    depending on the kernel matrix used in the code construction.  The
    task of determining $\beta$ for each kernel matrix has been fully
    resolved; interested readers are referred to \cite{ArT09, KSU10,
    HMT13, MoT14}.

    On the other branch, called the \emph{scaling exponent} branch, $f$
    is a constant\footnote{Sometimes $f$ is not a constant but
    converging to $0$.  For instance $f(n) = \exp(-n^{2/3})$.  In this
    case, $2^n f(n)$, the upper bound on the miscommunication
    probability, will exceed $1$, so the corresponding polar code is
    useless.  Yet the asymptotics of $g$ still helps us study other
    useful codes.} and the asymptotics of $g$ is examined.  For BECs,
    \cite{HAU10, KMT10} managed to estimate that $g(n) \approx
    2^{-n/3.627}$.  For binary memoryless symmetric (BMS) channels, it
    can be shown that $g(n) < 2^{-n/\mu}$ for some constant $0 < \mu
    < \infty$ (with a really good decay of error $f(n) =
    \exp(-2^{0.49n})$) \cite{GuX15}.  This makes polar codes the only
    code family that is known to achieve capacity at a speed polynomial
    in the block length.  Further estimates of $\mu$ include $3.553 <
    \mu$ \cite{GHU12}, $3.579 < \mu < 6$ \cite{HAU14}, $\mu < 5.702$
    \cite{GoB14}, $\mu < 4.714$ \cite{MHU16}, and very recently $\mu <
    4.63$ \cite{Sub4.7}.  Now that we know the $\mu$ for polar codes and
    the optimal value being $\mu \approx 2$ for random codes
    \cite{BKB04, Hay09, PPV10}, the discrepancy begs the question: Can
    one modify polar codes to reach a smaller scaling exponent?

    The answer is positive:  Arıkan used the kernel matrix $\loll$ to
    combine and split channels.  Instead, one can use a larger matrix,
    for instance
    \begin{equation}
    \installlightgrayzero
        \left[ \begin{smallmatrix}
            1 & 0 & 0 & 0 & 0 & 0 & 0 & 0 \\
            1 & 1 & 0 & 0 & 0 & 0 & 0 & 0 \\
            1 & 0 & 1 & 0 & 0 & 0 & 0 & 0 \\
            1 & 0 & 0 & 1 & 0 & 0 & 0 & 0 \\
            1 & 1 & 1 & 0 & 1 & 0 & 0 & 0 \\
            1 & 1 & 0 & 1 & 0 & 1 & 0 & 0 \\
            1 & 0 & 1 & 1 & 0 & 0 & 1 & 0 \\
            1 & 1 & 1 & 1 & 1 & 1 & 1 & 1 \\
        \end{smallmatrix} \right]
        \label{mat:8x8}
    \end{equation}
    to combine and split channels.  In \cite{FaV14, YFV19, TrT21,
    Tro21s, BBL20, Lin21}, binary matrices ranging from $3 \times 3$ to
    $64 \times 64$ are studied and the scaling exponents over BECs are
    estimated.  The best scaling exponent up to every matrix size is
    plotted in \cref{fig:large}.  There are also meta-asymptotic results
    stating that $\mu \approx 2$ can be achieved using larger and larger
    matrices.  This statement was proved over $q$-ary erasure channels
    \cite{PfU16}, binary erasure channels \cite{FHM21}, all BMS channels
    \cite{GRY22}, and finally discrete memoryless channels
    \cite{Hypotenuse}.

    As much as we want to lower polar code's scaling exponent, there is
    one caveat that renders large matrices impractical:  the smallest
    matrix whose scaling exponent is strictly better than $\loll$ is the
    $8 \times 8$ matrix given in \eqref{mat:8x8}.  Using this matrix
    takes twice more time to decode (estimate based on the method of
    \cite{BFS17}), whereas the benefit we gain is that $\mu$ slightly
    decreases from $3.627$ to $3.577$.  As the matrix gets larger and
    deviates more from the tensor powers of $\loll$, the time complexity
    grows drastically.  For this reason, it is unlikely that we will
    ever see polar code based on large matrices (unless it is for other
    concerns \cite{BGL20}).

    Large matrices aside, many other techniques emerge with empirical
    evidence that they improve polar codes---concatenation, cyclic
    redundancy check, and list decoder to name a few.  But none of them
    sees a proof of improvements in the scaling exponent; in fact, quite
    the opposite was reported \cite{MHU15}.  So we are back to the
    drawing board where we want to improve polar codes' scaling exponent
    while minimizing the complexity penalty.

    One approach that seems promising, albeit very little is known, is
    to use a non-binary input alphabet.  This line of research dates
    back to low-density parity-check codes \cite{AnA16, RaU05}.  For
    polar code, it started from Şaşoğlu \cite{STA09, Sas12, Chi14},
    wherein the goal was to find at least one way to polarize arbitrary
    finite alphabets regardless of the speed.  In particular, the usual
    matrix $\loll$ is known to polarize prime fields.  Later,
    Sahebi--Pradhan \cite{SaP13} and Park--Barg \cite{PaB13} showed that
    $\loll$ cannot polarize non-prime fields.  Then, Mori--Tanaka
    \cite{MoT14} classified all matrices that can polarize finite fields
    (i.e., the alphabet size must be a prime power).  One step forward,
    Nasser \cite{Nas16} classified all binary operators (i.e., bivariate
    functions) that can polarize arbitrary finite alphabets.  In
    \cite{BGN22, BGS18}, the authors showed that, for any polarizing
    matrix over prime fields, one has $\mu < \infty$.  In
    \cite{Hypotenuse}, the authors showed that $\mu \approx 2$ is
    reachable over arbitrary finite alphabets.




    Why is a non-binary input alphabet attractive?  There are at least
    three reasons.  First, modulation: For quadrature amplitude
    modulation (QAM) and amplitude and phase-shift keying (APSK), a
    constellation point is more likely to be confused with constellation
    points nearer to it.  A non-binary channel models this proximity
    relation more naturally than a series of correlated binary channels
    do \cite{SSS13, CKW19}.  Second, two-stage polarization: If we
    weakly-polarize a binary channel with $\loll$, treat every two
    binary channels as one quaternary channel, and strongly-polarize the
    quaternary channels with the $4 \times 4$ Reed--Solomon matrix, we
    can improve the asymptotics of $f(n)$ from $\exp(-2^{0.5n})$ to
    $\exp(-2^{0.5731n})$ \cite{PSL16} (see also \cite{AMV22, CBM18}).
    Third, and most importantly, scaling exponent: Several works have
    observed that non-binary matrices of the form $\lowl$ just polarize
    faster than $\loll$ \cite{YuS18, LiY21, Sav21}; some reported that
    even a minuscule amount of permutation can achieve similar effects
    \cite{LYH23}.  Could it be that the non-binary scaling exponents are
    smaller?


    Consider \cite{RWL22}'s technique that uses $\loll$ to polarize
    non-binary channels; their result has an implication that non-binary
    channels' scaling exponent is at least as good as binary channels'.
    In this paper, we aim to answer the question of whether the former
    is strictly better than the latter.  By defining a toy model that
    contains a pair of BECs as a special case and estimating the scaling
    exponent of $\lowl$, we provide a proof of concept result that an
    expansion in alphabet size does result in an improvement in scaling
    exponent.  Recall that BECs form a one-parameter family and that
    this property makes its scaling behavior easy to analyze.  This
    paper's overall strategy is to show that the descendants of a
    quaternary channel converge to an almost--one-parameter family; we
    then analyze the scaling behavior of this family and conclude the
    following.

    \begin{theorem}[main theorem]
        Treating a pair of BECs as a quaternary channel, the $2 \times
        2$ matrix $\lowl$ over $\FF_4$ induces a scaling exponent less
        than $3.451$.  Here, $\omega^2 + \omega + 1 = 0$.
    \end{theorem}

    This paper is organized as follows.
    \Cref{sec:polar} reviews polar code.
    \Cref{sec:model} defines tetrahedral erasure channels (TECs) as a
    generalization of pairs of BECs, defines balanced TECs to be those
    that possess some symmetry, and defines edge-heavy TECs to be those
    that will be polarized faster.
    \Cref{sec:kernel} defines serial combination and parallel
    combination that will be used to polarize TECs.
    \Cref{sec:balance} shows that unbalanced TECs tend to become very
    close to balanced TECs, so it suffices to consider the speed of
    polarization of the latter.
    \Cref{sec:trap} shows that balanced TECs tend to become very close
    to edge-heavy TECs, so it suffices to consider the speed of
    polarization of the latter.
    \Cref{sec:main} estimates the speed of polarization of balanced
    edge-heavy TECs, which proves the main theorem.

\section{Polar Code}                                   \label{sec:polar}

    Readers who are familiar with polar code may skip this section.
    This section serves a high-level summary of polar code.  More
    details are found in \cite[Chapter~2]{Chilly}.  We assume BEC
    throughout the section.

    Let $X \in \FF_2$ be a random variable following the uniform
    distribution.   Let $Y \in \FF_2 \cup \{?\}$ be a random variable
    with transition probabilities
    \begin{align*}
        \PP\{\, Y = x \,|\, X = x \,\} & = 1 - \varepsilon, \\
        \PP\{\, Y = {?} \,|\, X = x \,\} & = \varepsilon.   
    \end{align*}
    Here, $\varepsilon \in [0,1]$ is called the \emph{erasure
    probability}.  The pair $(X|Y)$ is called a \emph{binary erasure
    channel} (BEC) and denoted by $\BEC(\varepsilon)$.  The conditional
    entropy $H(\BEC(\varepsilon)) = H(X|Y) = \varepsilon$ is defined
    through Shannon's mean.

    Let $(X_1|Y_1)$ and $(X_2|Y_2)$ be two iid copies of
    $\BEC(\varepsilon)$.  Define the serial combination
    $\BEC(\varepsilon)^\seri$ to be $( X_1+X_2 | Y_1,Y_2 )$.  That is,
    what do we know about $X_1 + X_2$ when given $Y_1$ and $Y_2$?  One
    sees that it is information theoretically equivalent to
    $\BEC(2\varepsilon-\varepsilon^2)$.  Define the parallel combination
    $\BEC(\varepsilon)^\para$ to be $( X_1 | Y_1,Y_2,X_1+X_2 )$.  That
    is, what do we know about $X_1$ when given $Y_1$, $Y_2$, and
    $X_1+X_2$?  We can see that it is information theoretically
    equivalent to $\BEC(\varepsilon^2)$.

    Serial and parallel combinations apply recursively.  A polar code of
    block length $2^n$ is specified by a subset of strings $\II
    \subseteq \{\seri,\para\}^n$.  In this code, a synthetic channel
    \begin{equation}
        \Bigl( \dotsb \bigl( ( \BEC(\varepsilon)
            ^{c_1} )^{c_2} \bigr) \dotsb \Bigr)^{c_n}
        \label{for:tower}
    \end{equation}
    will be used to transmit useful information iff $(c_1, c_1, \dotsc,
    c_n) \in \II$.  The code rate of this polar code is $|\II|/2^n$.
    The exact miscommunication probability of this polar code is hard to
    find, but has an upper bound of
    \[
        \sum_\II H \biggl( \Bigl( \dotsb \bigl( ( \BEC(\varepsilon)
            ^{c_1} )^{c_2} \bigr) \dotsb \Bigr)^{c_n} \biggr).
    \]

    To define a good $\II$, choose a function $f(n)$ and collect all
    strings $(c_1, c_1, \dotsc, c_n) \in \{\seri,\para\}^n$ such that
    $H(\text{channel~\eqref{for:tower}})$ is less than $f(n)$.  The fact
    that the erasure probabilities undergo simple evolutions
    $\varepsilon \mapsto 2\varepsilon-\varepsilon^2$ and $\varepsilon
    \mapsto \varepsilon^2$ motivates the following stochastic process:
    define $\{H_n\}_n$ by initial value $H_0 \coloneqq \varepsilon$ and
    evolution rule $H_{n+1} \coloneqq 2H_n - H_n^2$ or $H_n^2$ with
    equal probability.  Then the code rate $|\II|/2^n$ coincides with
    $\PP \{\, H_n \leq f(n) \,\}$.  The gap to capacity $g(n) \coloneqq
    1 - H_0 - |\II|/2^n = 1 - H_0 - \PP \{\, H_n \leq f(n) \,\}$ can
    also be expressed using $f$.

    In a way, the study of polar code over BEC is the study of the cdf
    of $H_n$, with emphasis put on the hard threshold at $1 - H_0$.
    Abusing the same logic, this paper is a study of a stochastic
    process $\{W_n\}_n$ that lives in the four-dimensional simplex
    $[0,1]^5 \cap \{\, p+q+r+s+t = 1 \,\}$, which happens to have
    implications in coding theory (the implication being how to use
    simple instruments to construct low-$\mu$ codes).

\section{An Inspirational Channel Model}               \label{sec:model}

    We are to define a type of quaternary channels in this section.
    This should be the smallest possible set of quaternary channels that
    meet the following two criteria: (a) it should model a pair of BECs
    as a special case; and (b) it should be closed under pre-processing
    the input using invertible linear transformations.  Allowing such
    pre-processing is crucial to the improvement of scaling exponent
    as it helps mixing the information.

\begin{figure*}\centering
    \iflabor
    \begin{tikzpicture}
        \foreach \x in{-1, 1}{
            \foreach \y in{-1, 1}{
                \draw [xscale=\x, yscale=\y] (0, 3.5) -| (4.4, 3.1) -|
                      (5, 3.6) -| (5.6, 3) -| (5.1, 2.5) -| (5.5, 0);
            }
        }
        \node at (0, -3.5) [anchor=base] {discrete memoryless channels};
        \draw (-5, -2.8) -- (-3.7, 3.3) -- (1.6, 3.1) -- (2.5, -3) --
              node [sloped, anchor=base] {quaternary channels} cycle;
        \draw (-3.5, -2.2) rectangle +(5, 5) (-1, -2.2)
              node [anchor=base] {tetrahedral erasure channels};
        \foreach \a in {0, 90, 180, -90} {
            \draw [shift={(0.5, 2)}, rotate=\a]
                  (-0.5, 0.7) -| (0.5, 0.5) -- (0.7, 0.5);
        }\node at (0.5, 2) [align=center] {BEC\\[-2pt]$\times$\\BEC};
        \draw (-3.4, -1.5) -- (-1, 2.7) -- (1.4, -1.5) --
              node [anchor=base] {balanced TECs} cycle;
        \draw (-3, -1) to [bend left=30]
              node [anchor=base] {edge-heavy TECs}
              (1, -1) to [bend right=60, looseness=2] cycle;
        \draw [postaction=decorate, decoration={text along path,
               text align=center, text={binary erasure channels}}]
              [line width=2pt] (3.5, -2.5) to[bend right=60] (3.5, 2.5);
        \draw (3.5, -2.5) to [bend left=60] (3.5, 2.5);
        \node at (3.5, 0) [align=center]
              {binary\\memoryless\\symmetric\\channels};
    \end{tikzpicture}
    \fi
    \caption{
        The Euler diagram of channels featured in this paper.  The cross
        is the set of pairs of BECs; it will converge to the set of
        balanced TECs (\cref{sec:balance}). The balanced TECs will then
        converge to edge-heavy TECs (\cref{sec:trap}).  And then
        edge-heavy TECs polarize faster than BECs.  Note that BECs are a
        one-parameter family of extreme BMS channels, hence the thick
        curve.
    }                                               \label{fig:channels}
\end{figure*}
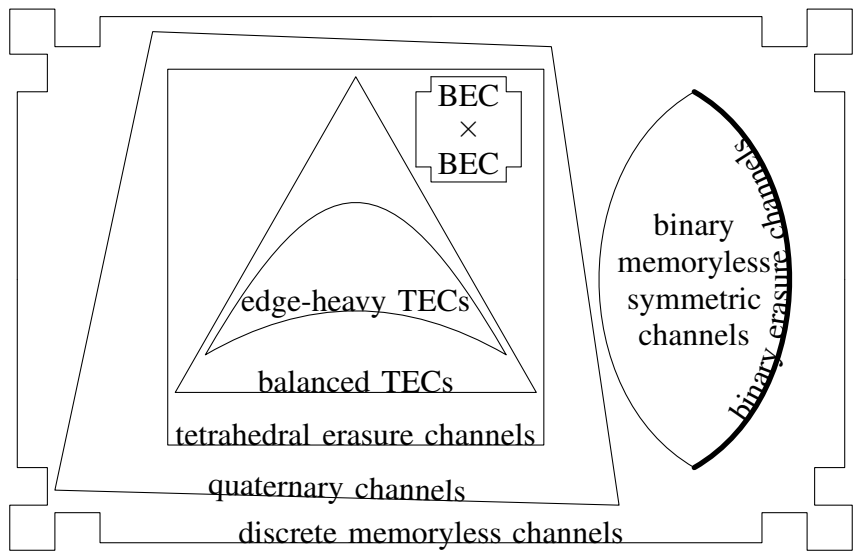

\subsection{Tetrahedral erasure channel}

    Let the input alphabet be $\FF_2^2$; and we assume the uniform input
    distribution throughout the paper.  For any input $(x_1,x_2) \in
    \FF_2^2$, the output will be in $(\FF_2 \cup \{?\})^3$ and assume
    one of the following five erasure patterns:
    \begin{itemize}
        \item $(x_1, x_1+x_2, x_2)$ with probability $p$;
        \item $(x_1,    ?   ,  ? )$ with probability $q$;
        \item $( ? , x_1+x_2,  ? )$ with probability $r$;
        \item $( ? ,    ?   , x_2)$ with probability $s$;
        \item $( ? ,    ?   ,  ? )$ with probability $t$.
    \end{itemize}
    We call $p$, $q$, $r$, $s$, and $t$ the \emph{subspace erasure
    probabilities} and they sum to $1$.  Such a channel is denoted by
    $\TEC(p, q, r, s, t)$.  For brevity, we say a TEC outputs $(x_1,
    x_2)$, outputs $x_1$, outputs $x_1 + x_2$, outputs $x_2$, and
    outputs nothing to represent the five erasure patterns.

    A TEC can be related to a tetrahedron whose vertices are at
    $(0,0,0)$\,,\, $(1,1,0)$\,,\, $(1,0,1)$\,,\, and $(0,1,1)$.
    Outputting $(x_1,x_2)$ corresponds to the vertex $(x_1, x_1+x_2,
    x_2)$.  Outputting $x_1$ corresponds to the edge $(x_1, x_1,
    0)$---$(x_1, 1-x_1, 1)$.  Outputting nothing corresponds to the
    tetrahedron per se.  That is to say, a TEC takes a vertex as an
    input and outputs the same vertex with probability $p$, outputs one
    of the edges attached to that vertex with probabilities $q$, $r$,
    and $s$, respectively, and output the entire tetrahedron with
    probability $t$.

    There is another way to interpret a TEC.  Consider $\FF_4$ and let
    $\omega$ be a primitive element therein.  A TEC takes $x \coloneqq
    x_1 + x_2 \omega \in \FF_4$ as an input and outputs $x$,
    $\tr(x/\omega)$, $\tr(\omega x)$, or $\tr(x)$ or nothing (with a
    flag that allows the receiver to distinguish these five cases) with
    probabilities $p$, $q$, $r$, $s$, and $t$, respectively.  Here,
    $\tr\colon \FF_4 \to \FF_2$ is the field trace.  It is the matrix
    trace if we use matrices $\oooo, \lool, \olll, \lllo \in
    \FF_2^{2\times2}$ to represent $0, 1, \omega, 1 + \omega \in \FF_4$.

    On top of the the fact that TECs are a natural family of erasure
    channels, they relate to other channels that have been discussed in
    literature.

    \begin{proposition}                                \label{pro:4-ary}
        The ``$q$-ary erasure channel with erasure probability
        $\varepsilon$'' \cite{MoT10, PfU16}, when $q = 4$, is a TEC of
        the form $\TEC(\, 1-\varepsilon\,,\, 0\,,\, 0\,,\, 0\,,\,
        \varepsilon \,)$.
    \end{proposition}
    
    \begin{proposition}                                 \label{pro:pair}
        When transmitting two bits $x_1$ and $x_2$ through
        $\BEC(\delta)$ and $\BEC(\varepsilon)$, respectively, the
        outputs can be simulated by $\TEC(\,
        (1-\delta)(1-\varepsilon)\,,\,\allowbreak
        (1-\delta)\varepsilon\,,\, 0,
        \delta(1-\varepsilon)\,,\,
        \delta\varepsilon
        \,)$.
    \end{proposition}

    Proofs of \cref{pro:4-ary,pro:pair} are omitted.  The propositions
    imply that any scaling exponent estimate for TEC immediately
    generalizes to $4$-ary erasure channels and quaternary channels that
    are pairs of BECs.  Note that transmitting information over a pair
    of $\BEC(\varepsilon)$ is the same matter as transmitting over
    $\BEC(\varepsilon)$.  Hence, an estimate of the scaling exponent
    over TECs implies the same estimate for BECs.

    As H. Pfister pointed out, TEC has been studied in the context of
    non-binary low-density parity-check code by, among others, Rathi and
    Urbanke \cite{RaU05}, who pointed out that the idea of using
    non-binary alphabet dates all the way back to Gallager himself.
    Later in the polar code context, TEC became a very natural toy model
    for $2$-user multiple access channels \cite[Section~VII-B]{NaT16},
    as kindly pointed out by one of the anonymous reviewers.  More
    recently, \cite{LYH23} (and its successor \cite{LYH22+}) studied an
    even more flexible generalization of polar coding that, sketchily
    speaking, uses permutations to shuffle bits and uses erasure
    patterns to study the performance.  When the alphabet is an
    arbitrary abelian group, one can study erasure channels whose
    erasure patterns are projections onto quotient groups \cite{SaP13,
    SIF20}, as kindly pointed out by one of the anonymous reviewers.

\subsection{Channel functionals}

    The \emph{conditional entropy} (sometimes \emph{entropy}) of a TEC
    $W = \TEC(p, q, r, s, t)$ is defined as
    \[ H(W) \coloneqq \frac {q + r + s} 2 + t. \]
    This definition is compatible with Shannon's definition of
    conditional entropy in the sense that we lose one out of two bits of
    information with probability $q + r + s$ and two out of two bits
    with probability $t$.  Clearly $0 \leq H(W) \leq 1$.

    We define the \emph{moment of inertia} of a TEC $W = \TEC(p, q, r,
    s, t)$ as
    \[
        A(W)
        \coloneqq (q - r)^2 + (r - s)^2 + (s - q)^2.
    \]
    Clearly $0 \leq A(W) \leq 3$.  When $q = r = s$, we say that this
    TEC is \emph{balanced}.  Put it another way, a TEC is balanced iff
    the edges of the tetrahedron weigh the same iff its moment of
    inertia vanishes.  See also the ``symmetric over the product''
    condition in \cite{CLM21} and the ``equidistance'' condition in
    \cite{STA09}.

    We define the \emph{edge mass} of a TEC $W = \TEC(p, q, r, s, t)$ as
    \[ E(W) \coloneqq q + r + s. \]
    Clearly $0 \leq E(W) \leq 1$.  It is not hard to see that $H$ and
    $E$ uniquely determine a balanced TEC by
    \begin{align}
        p & = 1 - H(W) - \frac{E(W)}2, \label{for:p(x,y)}\\
        q & = r = s = \frac{E(W)}3, \label{for:r(x,y)}\\
        t & = H(W) - \frac{E(W)}2. \label{for:t(x,y)}
    \end{align}
    This implies that we can use $(H, E)$ to parametrize balanced TECs.
    While we all agree that $H$ is a very important parameter of
    channels, we will see that the $E$ is also important as it measures
    the volatility of $H$: a higher $E(W)$ means a higher $H(W^\seri) -
    H(W)$ and a higher $H(W) - H(W^\para)$.

    We define the \emph{Q-index} of a TEC $W$ by
    \[ Q(W) \coloneqq \frac {E(W)} {H(W) (1 - H(W))} .\]
    Clearly, $0 \leq E(W) \leq 2 \min(H(W), 1 - H(W))$ and hence $0 \leq
    Q(W) \leq 4$.
    
    We call a TEC $W = \TEC(p, q, r, s, t)$ \emph{edge-positive} if $qrs
    > 0$.  Edge positivity implies that the descendants of $W$ are all
    edge-positive and have positive $E$, $H$, and $1 - H$, which implies
    that $Q$ is always well-defined.  We call a TEC $W$
    \emph{edge-heavy} if $Q(W) \geq \alpha \coloneqq 2\sqrt7 - 4$.  Note
    that, for a BEC $V$, $H(V^\seri) - H(V) = H(V) - H(V^\para) = H(V)
    (1 - H(V))$.  Hence we treat $H(1 - H)$ as the standard volatility.
    And then we compute the quotient of $E(W)$ by the standard value to
    get a normalized volatility measure.  Why we choose $\alpha$ as the
    threshold will be clear later.

\section{Channel Synthesis}                           \label{sec:kernel}

    TECs can be serially combined or parallelly combined as in the
    theory of density evolution \cite{LaH06}.
    
    Simply put, the serial combination of two channels is analogous to a
    standardized math exercise where, in order to test if students know
    both $u$ and $v$, we ask for $u + v$.

    Parallel combination, on the other hand, is analogous to a generous
    exercise where we give the true value of $u + v$ as a hint, and
    so any student who can compute $u$ or $v$ will immediately know
    both quantities.

    In this section, we demonstrate that the serial and parallel
    combinations of two TECs are again TECs, and we will study how the
    subspace erasure probabilities evolve under combinations.

\subsection{Serial combination}

    Let $U$ and $V$ be two independent channels with subspace erasure
    probabilities $\TEC(p, q,r, s, t)$ and $\TEC(p', q', r', s', t')$,
    respectively.  The \emph{serial combination} of $U$ and $V$ is
    defined to be the task of guessing $(u_1 + v_1, u_2 + v_2)$ given
    the output of inputting $(u_1, u_2)$ into $U$ and the output of
    inputting $(v_1, v_2)$ into $V$.  As $U$ produces five erasure
    patterns and $V$ also produces five, there are twenty-five erasure
    patterns in total.

    \begin{enumerate}
        \item With probability $pp'$, $U$ outputs $(u_1, u_2)$ and $V$
            outputs $(v_1, v_2)$:  In this case, we can infer both $u_1
            + v_1$ and $u_2 + v_2$.
        \item WIth probability $pq'$, $U$ outputs $(u_1, u_2)$ and $V$
            outputs $v_1$:  In this case, we can infer $u_1 + v_1$ but
            not $u_2 + v_2$.  Note that knowing $u_2$ does not reveal
            any information about $u_2 + v_2$ as $v_2$ is assumed to be
            uniformly randomly distributed.
        \item With probability $pr'$, $U$ outputs $(u_1, u_2)$ and $V$
            outputs $v_1 + v_2$:  In this case, we can not infer $u_1 +
            v_1$; nor can we infer $u_2 + v_2$.  However, we can still
            infer their sum $(u_1 + v_1) + (u_2 + v_2)$ because it is
            equal to $u_1$ (which is known) plus $u_2$ (which is also
            known) plus $v_1 + v_2$ (which is known as well).
        \item With probability $ps'$, $U$ outputs $(u_1, u_2)$ and $V$
            outputs $v_2$:  In this case, we cannot infer $u_1 + v_1$
            but we can infer $u_2 + v_2$.
        \item With probability $pt'$, $U$ outputs $(u_1, u_2)$ and $V$
            outputs nothing:  In this case, we obtain absolutely no
            information about $u_1 + v_1$ and $u_2 + v_2$ and their sum.
        
        \item With probability $qp'$, $U$ outputs $u_1$ and $V$ outputs
            $(v_1, v_2)$:  In this case, we can infer $u_1 + v_1$ but
            not $u_2 + v_2$.
    
        \item With probability $qq'$, $U$ outputs $u_1$ and $V$ outputs
            $v_1$:  In this case, we can infer $u_1 + v_1$ but not $u_2
            + v_2$.
        \item With probability $qr'$, $U$ outputs $u_1$ and $V$ outputs
            $v_1 + v_2$:  In this case, we know nothing about $u_1 +
            v_1$ because we cannot infer anything about $v_1$ given only
            $v_1 + v_2$.  We know nothing about $(u_1 + v_1) + (u_2 +
            v_2)$ because $u_2$ is missing.  Lastly, we know nothing
            about $u_2 + v_2$.  That is to say, we learn nothing useful.
        \item With probability $qs'$, $U$ outputs $u_1$ and $V$ outputs
            $v_2$:  In this case, we know nothing about $u_1 + v_1$ and
            $u_2 + v_2$ and their sum.
        \item  With probability $qt'$, $U$ outputs $u_1$ and $V$ outputs
            nothing:  In this case, we know nothing about $u_1 + v_1$
            and $u_2 + v_2$ and their sum.
        
        \item With probability $rp'$, $U$ outputs $u_1 + u_2$ and $V$
            outputs $(v_1, v_2)$:  In this case, we know $(u_1 + v_1) +
            (u_2 + v_2)$.
        \item With probability $rq'$, $U$ outputs $u_1 + u_2$ and $V$
            outputs $v_1$:  In this case, we learn nothing useful.
        \item With probability $rr'$, $U$ outputs $u_1 + u_2$ and $V$
            outputs $v_1 + v_2$:  In this case, we know $(u_1 + v_1) +
            (u_2 + v_2)$.
        \item With probability $rs'$, $U$ outputs $u_1 + u_2$ and $V$
            outputs $v_2$:  In this case, we learn nothing useful.
        \item With probability $rt'$, $U$ outputs $u_1 + u_2$ and $V$
            outputs nothing:  In this case, we learn nothing useful.
        
        \item With probability $sp'$, $U$ outputs $u_2$ and $V$
            outputs $(v_1, v_2)$:  In this case, we know $u_2 + v_2$.
        \item With probability $sq'$, $U$ outputs $u_2$ and $V$
            outputs $v_1$:  In this case, we learn nothing useful.
        \item With probability $sr'$, $U$ outputs $u_2$ and $V$
            outputs $v_1 + v_2$:  In this case, we learn nothing useful.
        \item With probability $ss'$, $U$ outputs $u_2$ and $V$
            outputs $v_2$:  In this case, we know $u_2 + v_2$
        \item With probability $st'$, $U$ outputs $u_2$ and $V$
            outputs nothing:  In this case, we learn nothing useful.

        \item With probability $tp'$, $U$ outputs nothing and $V$
        outputs $(v_1, v_2)$:  In this case, we learn nothing useful.
        \item With probability $tq'$, $U$ outputs nothing and $V$
        outputs $v_1$:  In this case, we learn nothing useful.
        \item With probability $tr'$, $U$ outputs nothing and $V$
        outputs $v_1 + v_2$:  In this case, we learn nothing useful.
        \item With probability $ts'$, $U$ outputs nothing and $V$
        outputs $v_2$:  In this case, we learn nothing useful.
        \item With probability $tt'$, $U$ outputs nothing and $V$
        outputs nothing:  In this case, we learn nothing useful.
    \end{enumerate}

\begin{table}\centering
    \def~{\textcolor{lightgray}}
    \begin{tabular}{c|ccccc}
        $\serial$ & P & Q & R & S & T \\[.2em]
        \hline\rule{0pt}{1.2em}%
        P         & P & Q & R & S &~T \\
        Q         & Q & Q &~T &~T &~T \\
        R         & R &~T & R &~T &~T \\
        S         & S &~T &~T & S &~T \\
        T         &~T &~T &~T &~T &~T \\
    \end{tabular}
    \hskip2em
    \begin{tabular}{c|ccccc}
        $\parall$ & P & Q & R & S & T \\[.2em]
        \hline\rule{0pt}{1.2em}%
        P         &~P &~P &~P &~P &~P \\
        Q         &~P & Q &~P &~P & Q \\
        R         &~P &~P & R &~P & R \\
        S         &~P &~P &~P & S & S \\
        T         &~P & Q & R & S & T \\
    \end{tabular}
    \vspace{1em}
    \caption{
        How erasure patterns evolve.  We use capital letter P to
        represent the erasure pattern whose probability is denoted by
        the lower letter $p$.  Same for Q, R, S, and T.
    }                                                \label{tab:combine}
\end{table}

    Case 1 is when we know both $u_1 + v_1$ and $u_2 + v_2$.
    Cases 2, 6, and 7 are when we know $u_1 + v_1$ but not the other.
    Cases 3, 11, and 13 are when we know the sum $(u_1 + v_1) + (u_2 +
    v_2)$ but not the summands.  Cases 4, 16, 19 are when we know $u_2 +
    v_2$ but not the other. Cases 5, 8, 9, 10, 12, 14, 15, 17, 18, 20,
    21, 22, 23, 24, and 25 are when we know nothing useful. Denote by $U
    \serial V$ the serial combination of $U$ and $V$; we conclude that
    it is a TEC with subspace erasure probabilities
    \begin{align*}
        U \serial V \coloneqq\TEC(
        & pp', \\
        & pq' + qq' + qp', \\
        & pr' + rr' + rp', \\
        & ps' + ss' + sp', \\
        & 1 - \text{the other four terms}
        ).
    \end{align*}
    See Table~\ref{tab:combine} for a summary.

\subsection{Parallel combination}

    The \emph{parallel combination} of $U$ and $V$ is
    defined to be the task
    of guessing $(u_1, u_2)$ given $(u_1 + v_1, u_2 + v_2)$ (the most
    informative output of $U \serial V$), the result of feeding $(u_1,
    u_2)$ into $U$, and the result of feeding $(v_1, v_2)$ into $V$.

    Denote by $U \parall V$ the parallel combination of $U$ and $V$.
    One can go over its twenty-five erasure patterns similar to what the
    previous subsection does.  For instance, with probability $qr'$, $U$
    outputs $u_1$ and $V$ outputs $v_1 + v_2$.  In this case, we can
    infer
    $v_1$ (using $u_1$ and $u_1 + v_1$), followed by
    $v_2$ (using $v_1$ and $v_1 + v_2$), and finally
    $u_2$ (using $v_2$ and $u_2 + v_2$);
    and hence we can completely recover $u_1$ and $u_2$.  Details
    omitted, it can be shown that $U \parall V$ is a TEC with subspace
    erasure probabilities
    \begin{align*}
        U \parall V \coloneqq \TEC(
        & 1 - \text{the other four terms}, \\
        & tq' + qq' + qt', \\
        & tr' + rr' + rt', \\
        & ts' + ss' + st', \\
        & tt'
        ).
    \end{align*}
    See also Table~\ref{tab:combine} for a summary.

    Note that there is a duality between $\TEC(p, q, r, s, t)$ and
    $\TEC(t, s, r, q, p)$ that respects $A$ and $E$, maps $H$ to $1 -
    H$, and swaps parallel and serial combinations:
    \begin{align*}
        \Dual( \TEC(p, q, r, s, t) ) & \coloneqq \TEC(t, s, r, q, p), \\
        \Dual( \Dual(W) )            & = W, \\
        H( \Dual(W) )                & = 1 - H(W), \\
        A( \Dual(W) )                & = A(W), \\
        E( \Dual(W) )                & = E(W), \\
        \Dual(U \serial V)           & = \Dual(U) \parall \Dual(V), \\
        \Dual(U \parall V)           & = \Dual(U) \serial \Dual(V).
    \end{align*}
    The duality grants us the convenience of proving half of a theorem
    because the other half follows by symmetry.

\subsection{Mori--Tanaka's twisting kernel}

    A $2 \times 2$ polarization \emph{kernel} $K$ over $\FF_4$ is
    defined with a ``twist'' as follows: For a pair of inputs $u, v \in
    \FF_4$, let $K$ be the linear transformation that reads $(u, v)
    \longmapsto (u+\omega v, v)$ or, equivalently,
    \[\begin{bmatrix}
        u & v
    \end{bmatrix} \longmapsto \begin{bmatrix}
        u & v
    \end{bmatrix}\begin{bmatrix}
        1 & 0 \\
        \omega & 1 \\
    \end{bmatrix}.\]
    This kernel was studied by Mori--Tanaka \cite{MoT14} and is shown to
    be polarizing.  If we treat $\FF_4$ as $\FF_2^2$ and express $u$ and
    $v$ as $u = u_1 + u_2 \omega$ and $v = v_1 + v_2 \omega$ for $u_1,
    u_2, v_1, v_2 \in \FF_2$, then $K$ reads
    $\bigl(\,
        (u_1,u_2)\,,\,
        (v_1,v_2)
    \,\bigr)
    \longmapsto
    \bigl(\,
        (u_1 + v_2, u_2 + v_1 + v_2)\,,\,\allowbreak
        (v_1,v_2)
    \,\bigr)$.
    Equivalently,
    \[ \begin{bmatrix}
        u_1 & u_2 & v_1 & v_2
    \end{bmatrix} \longmapsto \begin{bmatrix}
        u_1 & u_2 & v_1 & v_2
    \end{bmatrix}
    \installlightgrayzero 
    \left[ \begin{smallmatrix}
        1 & 0 & 0 & 0 \\
        0 & 1 & 0 & 0 \\
        0 & 1 & 1 & 0 \\
        1 & 1 & 0 & 1 \\
    \end{smallmatrix} \right]. \]
    The kernel $K$ combines two TECs $U$ and $V$ to synthesize $U
    \serial (V\omega)$ and $U \parall (V\omega)$, where $V\omega$ is the
    channel that multiplies the input by $\omega$ before feeding it into
    $V$.  For brevity, $W \serial (W\omega)$ and $W \parall (W\omega)$
    are denoted by $W^\seri$ and $W^\para$, respectively.

    Multiplying a TEC by $\omega$ behaves like a rotation of order $3$
    (after all, $\omega^3 = 1$ and it is rotating the tetrahedron).  It
    maps $\TEC(p,q,r,s,t)$ to $\TEC(p,s,q,r,t)$.  If $W$ is balanced,
    rotation does not alter it: $W = W\omega$.  If it is not balanced,
    then the rotation helps to break the alignment of $q$, $r$, and $s$
    so that a large subspace erasure probability is paired with a small
    one.  More precisely, in
    \begin{align*}
        \TEC(p, q, r, s, t) ^ \seri \coloneqq \TEC(
        & p^2, \\
        & ps + sq + qp, \\
        & pq + qr + rp, \\
        & pr + rs + sp, \\
        & 1 - \text{the other 4 terms}
        )
    \end{align*}
    and
    \begin{align*}
        \TEC(p, q, r, s, t) ^ \para \coloneqq \TEC(
        & 1 - \text{the other 4 terms}, \\
        & ts + sq + qt, \\
        & tq + qr + rt, \\
        & tr + rs + st, \\
        & t^2
        ),
    \end{align*}
    we see that the ``diagonal terms'' $qq'$, $rr'$, and $ss'$ are
    replaced by ``misalign terms'' $qr$, $rs$, and $sq$.  This hints
    that twisting makes it easier to reduce $q$, $r$, and $s$ by
    redistributing the masses to $p$ and $t$.
    
\subsection{Channel process}

    For a TEC $W$, we call $W^\seri$ the \emph{serial-child} of $W$ and
    $W^\para$ the \emph{parallel-child} of $W$.  Together, they are the
    \emph{children} of $W$.  The \emph{descendants} of $W$ are the
    children of $W$ together with the descendants of the children of
    $W$.  The \emph{$n$th-generation} descendants of $W$ are the
    $(n-1)$th-generation descendants of the children of $W$; the $0$th
    is $W$ itself.

    When $W$ is understood from the context, let $W_0$ be $W$.  For $n$
    a positive integer, let $W_n$ be a random child of $W_{n-1}$ with
    equal probability.

    The common strategy used to estimate the scaling exponent concerns a
    concave function $\psi\colon [0,1] \to \RR$ such that $\psi(0) =
    \psi(1) = 0$ and it is positive elsewhere.  With $\psi$, one finds a
    $0 < \mu < \infty$ such that
    \[
        \frac {\psi(H(W^\seri)) + \psi(H(W^\para))} {2\psi(H(W))}
        \leq 2^{-1/\mu}.
    \]
    We will treat the preceding formula as an operational definition of
    the scaling exponent because, with this formula, a routine argument
    \cite[Sections 5.8--5.10]{Chilly} will show that
    \[
        \PP \bigl\{\, H(W_n) < \exp \bigl( -e^{n^{1/3}} \bigr) \,\bigr\}
        > 1 - H(W_0) - 2^{-n/\mu}.
    \]
    That is to say, a good choice of $\psi$ provides a good
    characterization of the scaling behavior of polar codes.

\section{Unbalanced TEC Becomes Balanced}            \label{sec:balance}

    In this section, we argue that TECs undergoing the polarization
    process tend to become more balanced than before.  We do so by
    showing that the moments of inertia are decreasing.  We begin
    by showing that $A(W^\seri) + A(W^\para) \leq A(W)$, which implies
    that the total amount of moments of inertia is conserved.

    \begin{proposition}[conservation of inertia]  \label{pro:conserve-A}
        $A(W^\seri) + A(W^\para) \leq A(W)$ for any TEC $W$.
    \end{proposition}

    A proof of \cref{pro:conserve-A} is in \cref{pf:conserved-A}.  By
    the proposition, $A(W) \geq A(W^\seri) + A(W^\para) \geq
    A(W^{\seri\seri}) + A(W^{\seri\para}) + A(W^{\para\seri}) +
    A(W^{\para\para}) \geq A(W^{\seri\seri\seri}) \cdots$. Hence the
    expectation of $A(W_n)$ over all $W_n$ is at most $A(W)/2^n$.  By
    the same idea behind Markov's inequality, this suggests that for any
    TEC $W$, all but a few descendants have exponentially small
    $A$'s: they are almost-balanced.

    However, \cref{pro:conserve-A} does not completely rule out the
    case that some descendants will be highly unbalanced and eventually
    slow down the polarization.  So we present a uniform control on $A$.

    \begin{proposition}[uniform loss of inertia]   \label{pro:uniform-A}
        $A(W^\seri)$, $A(W^\para) \leq A(W) (1 - A(W)/3)$ for any TEC
        $W$.
    \end{proposition}

    A proof of \cref{pro:uniform-A} is in \cref{pf:uniform-A}.  Now the
    recurrence relation $A(W_{n+1}) \leq A(W_n) \* (1 - A(W_n)/3)$ is
    equivalent to $A(W_{n+1}) - A(W_n) \leq - A(W_n)^2/3$ and analogous
    to the ordinary differential equation $f'(t) \leq - f(t)^2/3$.
    Solving it, we get $f(t) = O(1/t)$.  Hence we expect that
    $A(W_n) = O(1/n)$.

    \begin{corollary}[ultimate loss of inertia]   \label{cor:ultimate-A}
        Fix a TEC $W$, then $A(W_n) = O(1/n)$ as $n \to \infty$.
    \end{corollary}
    
    \begin{IEEEproof}
        We prove by induction that $1/A(W_n) \geq n/3$.  Base case:
        By the definition of moment of inertia, $A(W_1) \leq 3$, hence
        $1/A(W_n) \geq 1/3$.
        Induction case:
        \begin{align*}
            \frac 1 {A(W_{n+1})}
            & \geq \frac 1 {A(W_n)(1 - A(W_n)/3)} \\
            & \geq \frac {1 + A(W_n)/3} {A(W_n)} \\
            & = \frac 1 {A(W_n)} + \frac13 \\
            & \geq \frac n3 + \frac13.
        \end{align*}
        This\footnote{ This elegant proof is provided by one of the
        anonymous reviewers.}
        finishes the induction and the proof that $A(W_n) = O(1/n)$.
    \end{IEEEproof}

    From \cref{pro:conserve-A} to \cref{cor:ultimate-A}, these results
    all lead to the same conclusion that any unbalanced TEC will quickly
    become very similar to a balanced one. Therefore, we expect that the
    speed of polarization of unbalanced TECs is dominated by that of
    balanced TECs.

    We invite readers to assume that it suffices to consider balanced
    TECs when estimating the scaling exponent over TECs.  These readers
    may jump to the next section, where we will be studying the
    evolution of $\{H(W_n)\}_n$ and $\{E(W_n)\}_n$ for balanced $W$'s.
    For readers who want to see a complete, rigorous proof, we introduce
    the following technical results that will be used later.

    \begin{proposition}[continuity in inertia]  \label{pro:continuous-A}
        For any TEC $W = \TEC(p, q, r, s, t)$ and its balanced version
        \[
            \bar W \coloneqq \TEC \Bigl(
                p,
                \frac{q + r + s}{3},
                \frac{q + r + s}{3},
                \frac{q + r + s}{3},
                t
            \Bigr).
        \]
        we have
        \begin{align*}
            H(W)       & = H(\bar W), \\
            E(W)       & = E(\bar W), \\
            H(W^\seri) & = H(\bar W^\seri) + \frac{A(W)}{12}, \\
            E(W^\seri) & = E(\bar W^\seri) - \frac{A(W)}{6}, \\
            H(W^\para) & = H(\bar W^\para) - \frac{A(W)}{12}, \\
            E(W^\para) & = E(\bar W^\para) - \frac{A(W)}{6}.
        \end{align*}
    \end{proposition}


    \begin{theorem}[monotonicity of $A/E$]      \label{thm:monotone-A/E}
        For any TEC $W$,
        \[
            \frac{A(W^\seri)}{E(W^\seri)}, \frac{A(W^\para)}{E(W^\para)}
            \leq \frac{A(W)}{E(W)}
            \leq 2E(W).
        \]
    \end{theorem}

    \begin{theorem}[fast loss of inertia]             \label{thm:fast-A}
        For any TEC $W$,
        \begin{align*}
            A(W^\para) & \leq A(W) H(W)^2, \\
            A(W^\seri) & \leq A(W) (1 - H(W))^2.
        \end{align*}
    \end{theorem}

    A proof of \cref{pro:continuous-A} is in \cref{pf:continuous-A}.  A
    proof of \cref{thm:monotone-A/E} is in \cref{pf:monotone-A/E}.  A
    proof of \cref{thm:fast-A} is in \cref{pf:fast-A}.  Note that
    \cref{thm:fast-A} generalizes \cref{pro:conserve-A} as $H(W)^2 + (1
    - H(W))^2 \leq 1$.

\section{Balanced TECs Hoard Edge Mass}                 \label{sec:trap}

    In this section, we want to show that the Q-index $Q(W_n) = E(W_n) /
    H(W_n) (1 - H(W_n))$ of a sufficiently deep descendant is about
    $1.6$.  Put another way, there is a ``trap'' that constrains the
    relation between $E(W_n)$ and $H(W_n)$.  Our main strategy is to
    show that, if $Q(W_n)$ is too low, $Q(W_{n+1})$ will become higher,
    and vice versa.

    Recall that $W = \TEC(p, q, r, s, t)$ is said to be edge-positive if
    $qrs > 0$, which implies that $E, H, 1 - H$ are all positive for all
    descendants of $W$, which then implies that $Q$ are always defined.
    Recall also that $W$ is said to be edge-heavy if $Q(W) \geq \alpha
    \coloneqq 2\sqrt7 - 4$.  Note that $\alpha \approx 1.3$.

    \begin{theorem}[trapping region]                  \label{thm:trap-Q}
        If $W$ is balanced and edge-heavy, then its children are
        balanced and edge-heavy.
    \end{theorem}

    A proof of \cref{thm:trap-Q} is in \cref{pf:trap-Q}. The theorem
    implies that all descendants of an edge-heavy TEC are edge-heavy.
    For a TEC that is not edge-heavy, its descendants will become
    ``edge-heavier'' by the following lemma.

    \begin{lemma}[attraction toward the trap]      \label{lem:attract-Q}
        Fix any $\varepsilon > 0$; choose $\delta \coloneqq
        3\varepsilon/8$.  Let $W$ be balanced and edge-positive.  We
        have that $Q(W) \leq \alpha - \varepsilon$ implies
        \begin{align*}
            Q(W^\seri) & \geq Q(W) \bigl( 1  + H(W) \delta \bigr), \\
            Q(W^\para) & \geq Q(W) \bigl( 1 + (1 - H(W)) \delta \bigr).
        \end{align*}
    \end{lemma}

    A proof of \cref{lem:attract-Q} is in \cref{pf:attract-Q}.  It is
    unfortunate that the factors $H(W)$ and $1 - H(W)$ before $\delta$
    slow down the rate at which $Q(W_n)$ approaches $\alpha = 2\sqrt7 -
    4$, especially when $H(W)$ is close to $0$ or $1$, respectively.
    These factors cannot be optimized away.  To see why, suppose that
    $H(W) = x \approx 1$ and $E(W) = y \approx 0$.  Then $H(W^\para)$ is
    about $x^2 + O(y^2)$ and $E(W^\para)$ is about $2xy + O(y^2)$.
    Hence $Q(W^\para)$ is about $2xy / x^2(1-x^2) \approx y / x(1-x) =
    Q(W)$.  That being the case, we can see from this example that TECs
    whose Q-indices can hardly be improved are already polarized, so we
    shall not worry about them.  Besides, we can prove uniform
    attraction.

    \begin{theorem}[uniform attraction]            \label{thm:uniform-Q}
        Fix any $\varepsilon > 0$.  For any balanced and edge-positive
        TEC $W$ such that $Q(W) \leq \alpha - \varepsilon$, there exists
        an integer $m > 0$ such that $Q(W_n) \geq Q(W) (1 +
        \varepsilon/8)$ for all $n \geq m$.
    \end{theorem}

    A proof of \cref{thm:uniform-Q} is in \cref{pf:uniform-Q}.  Uniform
    attraction means that every child is at least making some positive
    progress toward the trap.  Small steps of the descendants accumulate
    to a giant leap of the family.

    \begin{corollary}[ultimate attraction]        \label{cor:ultimate-Q}
        For any $\varepsilon > 0$ and any balanced and edge-positive TEC
        $W$, there exists an integer $m > 0$ such
        that $Q(W_n) \geq \alpha - \varepsilon$ for all $n \geq m$.
    \end{corollary}

    \begin{IEEEproof}
        Apply uniform attraction (\cref{thm:uniform-Q}) repeatedly.
        Every application improves the Q-index by a factor of $1 +
        \varepsilon / 8$.  So after a finite number of applications the
        Q-index can be made $\geq \alpha - \varepsilon$.
    \end{IEEEproof}
    
    To summarize this and the previous section, we have two trends:
    unbalanced TECs tend to become balanced; and ``edge-light'' TECs
    tend to become edge-heavy.
    
    The following proposition is a bound on the Q-indices in the
    opposite direction.
    
    \begin{proposition}[trap on the other side]      \label{pro:outer-Q}
        Let $W$ be a balanced, edge-positive TEC with $Q(W) \leq 2$.
        Then $Q(W^\seri) \leq 2$ and $Q(W^\para) \leq 2$.
    \end{proposition}
    
    A proof of \cref{pro:outer-Q} is in \cref{pf:outer-Q}.  This
    proposition is not required for the proof of the main theorem, but
    it helps us understand the dynamic of TECs.  Moreover, we anticipate
    that the counterparts of \cref{lem:attract-Q}, \cref{thm:uniform-Q},
    and \cref{cor:ultimate-Q} can all be stated and proved for bounding
    $Q$ from above.

    The following proposition gives a tighter (but more complex)
    trapping region than \cref{thm:trap-Q,pro:outer-Q} do.  A proof is
    omitted but the ideas are the same.  But even these are not the
    optimal trapping region.  For the optimal one, see the discussion in
    \cref{app:iterate}.

    \begin{proposition}[tighter trap]
        Let $f(x) \coloneqq x(1-x) (1.66 - 0.38x(1-x))$.  Then $E(W)
        \leq f(H(W))$ implies $E(W^\seri) \leq f(H(W^\seri))$ and
        $E(W^\para) \leq f(H(W^\para))$.  Let $g(x) \coloneqq x(1-x) (2
        - 2x(1-x)/3)$.  Then $E(W) \geq g(H(W))$ implies $E(W^\seri)
        \geq g(H(W^\seri))$ and $E(W^\para) \geq g(H(W^\para))$.
    \end{proposition}

    At the end of \cref{sec:balance}, we argue that since TECs will
    become practically indistinguishable from balanced TECs, all we need
    is to compute the scaling exponent for balanced TECs.  To convert
    that idea to rigorous proofs, one way is to generalize
    \cref{thm:trap-Q}--\cref{cor:ultimate-Q} to the case where $W$ is
    unbalanced.  We begin with the counterpart of \cref{thm:trap-Q}.

    \begin{theorem}[weak trap for the unbalanced]     \label{thm:weak-Q}
        There exists a small number $\varepsilon > 0$ such that, for any
        edge-positive $W$ (note: not necessarily balanced), we have
        $Q(W^\seri), Q(W^\para) \geq \min(Q(W), \varepsilon)$.
    \end{theorem}

    A proof of \cref{thm:weak-Q} is in \cref{pf:weak-Q}.  This theorem
    is the one of the two puzzle pieces of the following (somewhat
    topological) argument that addresses how to handle almost but not
    strictly-speaking balanced TECs.

    Suppose that the process $\{H(W_n)\}_n$ always stays in the interval
    $[a, d]$ for some constants $0 < a < d < 1$ such that $d - a \approx
    1$, then $H(W_n) (1 - H(W_n)) \geq a (1 - d)$ is bounded from below.
    Since $Q(W_n)$ is bounded from below by $\min(Q(W_0), \varepsilon)$
    by \cref{thm:weak-Q}, $E(W_n) = Q(W_n) H(W_n) (1 - H(W_n))$ is also
    bounded from below by $a (1 - d) \min(Q(W_0), \varepsilon)$.  Now we
    apply \cref{pro:continuous-A} to $W_n$ and its balanced version
    $\overline{W_n}$.  This allows us to predict $H(W_{n+1})$ and
    $E(W_{n+1})$ by
    \begin{align*}
        H(W_n^\seri) & = H(\overline{W_n}^\seri) + \frac{A(W_n)}{12}, \\
        E(W_n^\seri) & =  E(\overline{W_n}^\seri) - \frac{A(W_n)}{6}, \\
        H(W_n^\para) & = H(\overline{W_n}^\para) - \frac{A(W_n)}{12}, \\
        E(W_n^\para) & =  E(\overline{W_n}^\para) - \frac{A(W_n)}{6}.   
    \end{align*}
    But by \cref{cor:ultimate-A}, $A(W_n)$ will keep decreasing as $n
    \to \infty$ to the point that it becomes negligible compared to
    $H(W_{n+1})$, $1 - H(W_{n+1})$, and $E(W_{n+1})$ (which we know are
    bounded from below).  At that point, the additive error $O(A(W_n))$
    can be seen as a multiplicative error of $1 \pm (\text{some small
    number})$.  This means that we can \emph{apply}
    \cref{lem:attract-Q}--\cref{cor:ultimate-Q} up to some
    multiplicative error to the case where $W$ is unbalanced, instead of
    having to re-prove them for the unbalanced case.

    On the other hand, suppose that the process $\{H(W_n)\}_n$ escapes
    the interval $[b, c]$ at $n = m$, where $0 < a < b < c < d < 1$ and
    $c - b \approx 1$, and never reenters $[b, c]$.  Then either
    $H(W_n)$ or $1 - H(W_n)$ is very small.  Without loss of generality,
    let us discuss the case that $H(W_n)$ is small, where we have
    \begin{align*}
        H(W_n^\para) & \leq O(H(W_n)^2), \\
        H(W_n^\seri) & \geq 2 H(W_n) - O(H(W_n)^2)
    \end{align*}
    for all $n \geq m$.  This is the classical \emph{square-or-double}
    behavior of Ar\i kan's martingale.  The precise factor behind the
    big-$O$ do not affect the estimate of the scaling exponent if $b$ is
    taken to be small enough.\footnote{ In fact, the square-or-double
    behavior is even compatible with the optimal scaling exponent $2$
    \cite{GRY22}} To summarize this case, we do not have $Q(W_n) \geq
    \alpha$ but we do not need it anyways.

    There is one possibility left, namely the process $\{H(W_n)\}_n$
    escapes the interval $[a, d]$, but reenters $[b, c]$ later.  In this
    case, we cannot apply the square-or-double argument because $H(W_n)
    (1 - H(W_n))$ is not eventually small; nor can we apply
    \cref{lem:attract-Q}--\cref{cor:ultimate-Q} because $H(W_n) (1 -
    H(W_n))$ is sometimes too small.  That being said, we claim that we
    can choose a tiny $b$ (and $c \coloneqq 1 - b$) and an $a$ that is
    very tiny compared to $b$ (and $d \coloneqq 1 - a$) such that when
    $H(W_n)$ reenters $[b, c]$, $A(W_n)$ will be negligibly small and
    $Q(W_n)$ extremely close to $2$.

    \begin{theorem}[becomes extreme and then mediocre]
                                                   \label{thm:reenter-H}
        Let $\gamma > 0$ be fixed.  It is possible to select $0 < a < b
        < c < d < 1$ such that, for any realization of $\{H(W_n)\}_n$
        that escapes $[a, d]$ and then reenters $[b, c]$, $A(W_n) /
        E(W_{n+1}) < \gamma$ and $Q(W_n) > 2 - \gamma$ at the $n$ where
        the reentrance takes place.
    \end{theorem}
    
    A proof of \cref{thm:reenter-H} is in \cref{pf:reenter-H}.  This
    theorem is the last puzzle pieces of the topological argument that
    addresses how to handle almost but not strictly-speaking balanced
    TECs.

    The topological argument presented between \cref{thm:weak-Q} and
    \cref{thm:reenter-H} will be our strategy to prove the main theorem,
    which is done in the next section.  While we, the authors, did our
    best to simplify the argument, it is understandable that some
    readers might find it pleasant to read.  For those readers, we
    prepared a coding technique that immediately turns any unbalanced
    TEC into a perfectly balanced one.  See \cref{app:immediate}.

\section{Edge-heavy TECs Polarize Faster}               \label{sec:main}
    
    Let $W$ be any balanced TEC with $H(W) = x$ and $E(W) = y$.  Recall
    that $x$ and $y$ uniquely determine a balanced TEC via
    \eqref{for:p(x,y)}, \eqref{for:r(x,y)}, and \eqref{for:t(x,y)},
    hence we can compute and see that $H(W^\seri) = 2x - x^2 + y^2/12$
    and $H(W^\para) = x^2 - y^2/12$
 
    Note that $H(W^\seri)$ is is increasing in $y$ and $H(W^\para)$
    decreasing in $y$.  The monotonicity has two applications.
    Application one: If we know too little to lower bound $Q(W)$, we
    will upper bound $H(W^\para)$ using $x^2$.  In this case, the speed
    of polarization is at least $\mu \approx 3.627$, the number
    associated to BECs.  Application two: If we know $Q(W) \geq \alpha$,
    we will upper bound $H(W^\para)$ using $x^2 - (\alpha x(1-x))^2 /
    12$.  This time, $H(W^\para)$ and $H(W^\seri)$ are more separated so
    the speed of polarization is strictly better than $\mu \approx
    3.627$.  Any positive $\alpha$, not necessarily $2\sqrt7 - 4$, can
    improve the scaling.  This is demonstrated by the following lemma
    that uses $9/7$ in place of $\alpha$.
    
    \begin{lemma}[eigenfunction and eigenvalue]        \label{lem:eigen}
    Let $\psi(x) \coloneqq (x(1-x))^{0.697} (5 - \sqrt{x(1-x)})$.  For
    balanced TECs with $Q(W) \geq 9/7$, \[ \frac {\psi(H(W^\seri)) +
    \psi(H(W^\para))} {2\psi(H(W))} < 0.818.  \] \end{lemma}
    
    Comments on how to verify the lemma are in \cref{app:eigen}.  We are
    now ready for the main theorem.

    \begin{theorem}[main theorem]                       \label{thm:main}
        Consider a pair of BECs treated as a TEC, or consider any
        edge-positive TEC.  The $2 \times 2$ matrix $\lowl$ over $\FF_4$
        induces a scaling exponent less than $3.451$.
    \end{theorem}
 
    \begin{IEEEproof}
        Two iid copies of $\BEC(\varepsilon)$ can be seen as $W
        \coloneqq \TEC(\, (1-\varepsilon)^2\,,\,
        (1-\varepsilon)\varepsilon\,,\, 0,
        \varepsilon(1-\varepsilon)\,,\, \varepsilon^2 \,)$.  If
        $\varepsilon$ is $0$ or $1$, there is nothing to prove.  Suppose
        $0 < \varepsilon < 1$, then both $W^\seri$ and $W^\para$ have
        five positive subspace erasure probabilities.  (That is, their
        ``$p, q, r, s, t$'' are all positive).  They are edge-positive
        and so are the descendants.  Hence, the Q-index is always
        well-defined.

        By \cref{thm:weak-Q} and \cref{thm:reenter-H} and the discussion
        in between, we know that the behavior process $\{H(W_n)\}_n$ can
        be classified in to three categories (A), (B), and (C).

        (A) $H(W_n) \in [a, d]$ for all $n$ for some carefully selected
        $0 < a < d < 1$.  In this case, $A(W_n)$ converges to $0$
        while, thanks to \cref{thm:weak-Q}, $H(W_n)$, $1 - H(W_n)$,
        $Q(W_n)$, and $E(W_n)$ are all bounded from below.  Therefore,
        \cref{cor:ultimate-Q} applies.  We infer that $W_n$ are
        almost-balanced and almost--edge-heavy for all $n \geq m$ for
        some large number $m$ depending on $a$, $d$, and $W$.  In
        particular, we will have $Q(W_n) \geq 9/7$ for all $n \geq m$
        (because $9/7 \approx 1.286 < 1.291 \approx 2\sqrt7 - 4$); so
        the eigenvalues of the form
        \[
            \frac {\psi(H(W_n^\seri)) + \psi(H(W_n^\para))}
                  {2\psi(H(W_n))}
        \]
        are less than $0.818 < 2^{-1/3.451}$, where the value is
        estimated in \cref{lem:eigen}. 

        (B) $H(W_n) \in [0, 1] \setminus [b, c]$ for some carefully
        selected $0 < a < b < c < d < 1$ after $n = m'$ ($m'$ might
        be greater or less than $m$; it does not matter).  In this case,
        the $H$'s are in the \emph{square-or-double} region where,
        regardless of $A$ and $Q$,
        \[
            \psi(H(W_n^\seri))
            \approx \psi(2H(W_n) - O(H(W_n)^2))
            \approx \psi(2H(W_n))
        \]
        and
        \[
            \psi(H(W_n^\para))
            \approx \psi(O(H(W_n)^2))
            \approx 0
        \]
        Hence the eigenvalues are still less than $0.818 <
        2^{-1/3.451}$.  Before $n = \min(m, m')$, the eigenvalues can
        be upper bounded by $1$ due to the convexity of $\psi$.  Since
        only $\min(m, m')$ generations assume the bad eigenvalue, the
        overall scaling exponent is dominated by the improved value
        $3.451$.
        
        (C) For any realization of $\{W_n\}_n$ that does not fall into
        categories (A) and (B), it must be the case that $H(W_n)$, at
        some $n = m'$, escapes $[a, d]$ but reenters $[b, c]$ at some
        greater $n = m''$.  Before $n = m''$, category (B)'s argument
        applies: the eigenvalue can be upper bounded by  the improved
        value $0.818 < 2^{-1/3.451}$ for all but $\min(m, m')$
        generations.  So it remains to study what happens after $H(W_n)$
        reenters $[b, c]$.  By the discussion from \cref{thm:weak-Q} to
        \cref{thm:reenter-H}, we can select $a, b, c, d$ very carefully
        such that it takes sufficiently many generations to escape and
        reenter.  When the process is doing that, we see that it
        undergoes an unbalanced version of the square-or-double
        behavior, which makes it such that when $H(W_n)$ reenters $[b,
        c]$, $A(W_n)$ is low enough that we can pretend that $W_n$ is
        balanced and $Q(W_n)$ is high enough to enjoy the good scaling
        exponent.  The process $\{W_n\}_n$ may escape $[a, d]$ and
        reenter $[b, c]$ multiple times; but the same argument applies
        every time: When $H(W_n)$ is outside $[b, c]$, it assumes the
        good eigenvalue for the same reason as category (B).  When
        $H(W_n)$ is in $[b, c]$, it assumes the good eigenvalue for the
        same reason as category (A).  The key idea is that, every time
        $H(W_n)$ reenters $[b, c]$ from outside $[a, d]$, it does not
        have to wait for $m$ generations to raise its $Q$; it starts
        enjoying the good eigenvalue right away by \cref{thm:reenter-H}.

        From what we have discussed above, only the first $m$
        generations will assume the bad eigenvalue and all other
        generations will assume the good eigenvalue.  By considering
        polar code whose block length is large enough, the geometric
        mean of the eigenvalues approaches the good one.  Hence $W$, and
        hence any BEC, enjoys a scaling exponent less than $3.451$.
    \end{IEEEproof}

    In the abstract, we claim that the scaling exponent of $\lowl$ over
    TECs (and hence BECs) is $< 3.328$.  This number will be derived in
    \cref{app:iterate} with more intense numerical calculations.  In
    particular, there is a new trapping region that is bounded by two
    linear splines and is significantly smaller than the region bounded
    by $a x (1 - x)$ for $a = 2\sqrt7 - 4$ and $2$; the attraction
    toward the new trap is witnessed by sampling TECs with low
    edge-mass.  In \cref{app:simulate}, we also examine the actual
    values of $H(W_n)$ and its asymptotic behavior aligns with the
    estimate $3.328$.

\section{Conclusions}

    In this paper, we argue that $\lowl$ polarizes BECs faster than
    $\loll$ does.  We first show that a pair of BECs will be transformed
    into balanced TECs.  We then show that balanced TECs will be
    transformed into edge-heavy TECs.  Finally, we show that edge-heavy
    TECs assume a better scaling exponent.

\begin{figure}\centering
    \iflabor
    \begin{tikzpicture}
        \begin{axis}
            \addplot[PMS3015,mark=*,mark size=1pt,const plot]table{
                2	3.627
                8	3.577
                14  3.485
                16	3.346
                24	3.308
                26	3.301
                28	3.279
                30	3.272
                31	3.264
                32	3.122
                64	2.87
            };
            \draw[PMS1245](-9,3.451)--(99,3.451)(-9,3.328)--(99,3.328);	
        \end{axis}
    \end{tikzpicture}
    \fi
    \caption{
        Horizontal axis: matrix size;  vertical axis: scaling exponent
        of the best known matrix \cite{FaV14, YFV19, TrT21, Tro21s,
        BBL20}.  A matrix size will be skipped if no known matrix
        outruns all smaller matrices.  Underlying channel is a BEC.  Our
        estimates $3.451$ and $3.328$ are marked as yellow, horizontal
        lines.
    }                                                  \label{fig:large}
\end{figure}
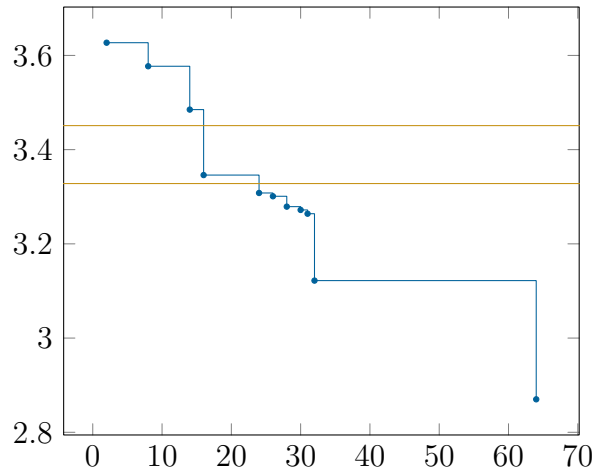

    Our rigorous overestimate of the scaling exponent is $3.451$; there
    is another overestimate of $3.328$ with strong numerical evidence.
    Compared to Arıkan's $2 \times 2$ matrix with $\mu \approx 3.627$,
    Fazeli--Vardy's $8 \times 8$ matrix with $\mu \approx 3.577$
    \cite{FaV14}, Trofimiuk--Trifonov's $16 \times 16$ matrix with $\mu
    \approx 3.346$ \cite{TrT21}, and Yao--Fazeli--Vardy's $32 \times 32$
    matrix with $\mu \approx 3.122$ \cite{YFV19}, our result suggests
    that one should consider expanding the alphabet size prior to
    enlarging the matrix size.  More precisely, the rigorous estimate is
    analogous to a $15 \times 15$ binary matrix; the more accurate
    estimate is analogous to a $20 \times 20$ binary matrix (see
    \cref{fig:large}).

\appendices
\crefalias{section}{appendix}
\crefalias{subsection}{appendix}

\section{Proofs for Balancing Channels}              \label{app:balance}

    For any TEC $W = \TEC(p,q,r,s,t)$,
    \begin{align*}
        A(W)       & = (q - r)^2 +(r - s)^2 +(s - q)^2, \\
        A(W^\seri) & = (q - r)^2 (s + p)^2 \\
                   &\, + (r - s)^2 (q + p)^2 \\
                   &\, + (s - q)^2 (r + p)^2, \\
        A(W^\para) & = (q - r)^2 (s + t)^2 \\
                   &\, + (r - s)^2 (q + t)^2 \\
                   &\, + (s - q)^2 (r + t)^2.   
    \end{align*}
    The first line is the definition and the two lines below are
    easily-verifiable algebraic identities.  They will be useful for
    this appendix.

\subsection{Average loss of inertia (Proposition~\ref{pro:conserve-A})}
                                                  \label{pf:conserved-A}

    We want to prove $A(W^\seri) + A(W^\para) \leq A(W)$.

    \begin{IEEEproof}
        First, note that $(q-r)^2 (r-s) (s-q) + (r-s)^2 (s-q) (q-r) +
        (s-q)^2 (q-r) (r-s)$ is zero by factoring out $(q-r) (r-s)
        (s-q)$.  Next, we want to prove that the following is
        nonnegative:
        \begin{align*}
            \kern0.5em&\kern-0.5em
            A(W) - A(W^\seri) - A(W^\para) \\
            &= \, (q - r)^2 \bigl( 1 - (s + p)^2 - (s + t)^2 \bigr) \\
            &\; + (r - s)^2 \bigl( 1 - (q + p)^2 - (q + t)^2 \bigr) \\
            &\; + (s - q)^2 \bigl( 1 - (r + p)^2 - (r + t)^2 \bigr) \\
            &= \, (q{-}r)^2
                  \bigl( 1-(s{+}p)^2-(s{+}t)^2-(r{-}s)(s{-}q) \bigr) \\
            &\; + (r{-}s)^2
                  \bigl( 1-(q{+}p)^2-(q{+}t)^2-(s{-}q)(q{-}r) \bigr) \\
            &\; + (s{-}q)^2
                  \bigl( 1-(r{+}p)^2-(r{+}t)^2-(q{-}r)(r{-}s) \bigr).  
        \end{align*}
        It remains to show that $1 - (s + p)^2 - (s + t)^2 - (r - s) (s
        - q)$ is nonnegative, as the rest will follow by symmetry.  To
        show so, replace $1$ with $(p + q + r + s + t)^2$.  One sees
        that $(p + q + r + s + t)^2 - (s + p)^2 - (s + t)^2 - (r - s)(s
        - q) = (2p + q + r + s + 2t)(q + r) + qr + 2pt$ is nonnegative.
    \end{IEEEproof}

\subsection{Uniform loss of inertia (Proposition \ref{pro:uniform-A})}
                                                    \label{pf:uniform-A}

    It suffices to prove $A(W^\para) \leq A(W) (1 - A(W)/3)$ for any TEC
    $W$ because $A(W^\seri) \leq A(W) (1 - A(W)/3)$ will follow by
    duality.

    \begin{IEEEproof}
        We have $(s + t)^2 = (1 - p - q - r)^2 \leq (1 - q - r)^2 \leq
        (1 - q + r) (1 + q - r) = 1 - (q - r)^2$.  So we see that the
        three terms that sum to $A(W^\para)$ can be bounded by
        \begin{align*}
            (q - r)^2 (s + t)^2 \leq (q - r)^2 (1 - (q - r)^2), \\
            (r - s)^2 (q + t)^2 \leq (r - s)^2 (1 - (r - s)^2), \\
            (s - q)^2 (r + t)^2 \leq (s - q)^2 (1 - (s - q)^2).   
        \end{align*}
        The average of these three terms is no greater than $(A(W)/3) (1
        - A(W)/3)$ because $A(W)/3$ is the average of $(q-r)^2$, $(r -
        s)^2$, and $(s - q)^2$ and because $x (1 - x)$ is concave in
        $x$.  We then conclude that $A(W^\para) \leq A(W) (1 - A(W)/3)$.
        The upper bound on $A(W^\seri)$ follows by duality.
    \end{IEEEproof}

\subsection{Continuity in inertia (Proposition~\ref{pro:continuous-A})}
                                                 \label{pf:continuous-A}

    Recall how $\bar W$, the balanced version of $W$, is defined: they
    have the same ``$p$'' and ``$t$'' but $\bar W$'s ``$q$'', ``$r$'',
    and ``$s$'' are set to be the average.  It is clear that $H(W) =
    H(\bar W)$ and  $E(W) = E(\bar W)$.  It remains to compare
    $E(W^\seri)$ and $E(\bar W^\seri)$ because the other identities will
    follow easily.

    \begin{IEEEproof}
        Compare
        \begin{align*}
            E(W^\para)
            & = sq + qr + rs + 2t (q + r + s), \\
            E(\bar W^\para)
            & = \frac {(q + r + s)^2} {3} + 2t (q + r + s).
        \end{align*}
        We infer that
        \begin{align*}
            \kern1em&\kern-1em
            E(\bar W^\para) - E(W^\para) \\
            & = \frac {2(q + r + s)^2 - (6sq + 6qr + 6rs)} {6} \\
            & = \frac{A(W)}{6}.
        \end{align*}
        For $H(\bar W^\para) - H(W^\para)$, notice that $H(W^\para) -
        E(W^\para)/2 = H(\bar W^\para) - E(\bar W^\para)/2 = t^2$
        and so the $t^2$ cancels.
    \end{IEEEproof}

\subsection{Monotonicity of \texorpdfstring{$A/E$}{A/E}
            (Theorem~\ref{thm:monotone-A/E})}    \label{pf:monotone-A/E}

    One way to relate $A$ to $E$ is via $A(W) = 2E(W)^2 - 6 (qr + rs +
    sq) \leq 2E(W)^2$, which proves the second inequality in the theorem
    statement.  We now want to show that $A(W^\seri) / E(W^\seri) \leq
    A(W) / E(W)$; this is enough for proving the first inequality in the
    theorem statement because $A(W^\para) / E(W^\para) \leq A(W) / E(W)$
    will also hold by symmetry.

    \begin{IEEEproof}
        It suffices to prove
        \[ A(W) E(W^\seri) (p+q+r+s+t) - A(W^\seri) E(W) \geq 0. \]
        For that, sum the following eleven terms that are obviously
        positive:
        \begin{gather*}
            p^2 (q+r+s) ( (q-r)^2 + (r-s)^2 + (s-q)^2 ), \\
            2 p ( (q^2 - r^2)^2 + (r^2 - s^2)^2 + (s^2 - q^2)^2 ), \\
            4 p ( qr (q-r)^2 + rs (r-s)^2 + sq (s-q)^2 ), \\
            3 p ( q^2 (r-s)^2 + r^2 (s-q)^2 + s^2 (q-r)^2 ), \\
            2 pt (q+r+s) ( (q-r)^2 + (r-s)^2 + (s-q)^2 ), \\
            2 t ( qr (q-r)^2 + rs (r-s)^2 + sq (s-q)^2 ), \\
            t ( q^2 (r-s)^2 + r^2 (s-q)^2 + s^2 (q-r)^2 ), \\
            2 qr (q+r) (q-r)^2, \\
            2 rs (r+s) (r-s)^2, \\
            2 sq (s+q) (s-q)^2 , \\
            2 qrs ( (q-r)^2 + (r-s)^2 + (s-q)^2 )
        \end{gather*}
        and we are done.  Remark: This is just Muirhead's inequality.
    \end{IEEEproof}

\subsection{Fast loss of inertia (Theorem~\ref{thm:fast-A})}
                                                       \label{pf:fast-A}
                                                         
    We want to show $A(W^\para) \leq A(W) H(W)^2$; this is enough for
    proving the theorem because $A(W^\seri) \leq A(W) (1 - H(W))^2$ will
    also hold by symmetry.

    \begin{IEEEproof}
        It suffices to demonstrate
        \[ A(W^\para) - A(W) H(W)^2 \geq 0. \]
        To do so, expand the left-hand side into a polynomial in $p, q,
        r, s, t$.  Then, observe that there is no monomial with $t^2$,
        $t^3$, or higher power.  The monomials that contain $t$ are
        \begin{gather*}
            {}
            + 2 q^3 t 
            + 2 r^3 t
            + 2 s^3 t
            {}
            + 6 q r s t \\
            {}
            - 2 q^2 r t
            - 2 q r^2 t
            - 2 r^2 s t
            - 2 r s^2 t
            - 2 s^2 q t
            - 2 s q^2 t
            .
        \end{gather*}
        We know that the sum of these monomials is nonnegative by
        Schur's inequality.  The monomials that do not contain $t$ are
        \begin{gather*}
            {}
            + q^4
            + r^4
            + s^4 \\
            {}
            + q^3 r
            + q r^3
            + r^3 s
            + r s^3
            + s^3 q
            + q^3 s \\
            {}
            + q^2 r s
            + q r^2 s
            + q r s^2 \\
            {}
            - 4 q^2 r^2
            - 4 r^2 s^2
            - 4 s^2 q^2
        \end{gather*}
        divided by $2$.  These monomials sum to
        \begin{gather*}
            (q + r + s) \cdot (\text{Schur's inequality}) \\
            {} + q r (q - r)^2 + r s (r - s)^2 + s q (s - q)^2,
        \end{gather*}
        which we can see is nonnegative.  Since the coefficients of
        $t^1$ and $t^0$ are nonnegative, the polynomial $A(W^\para) -
        A(W) H(W)^2$ is nonnegative.
    \end{IEEEproof}

\section{Proofs for Trapping Channels}                  \label{app:trap}

    Recall that, if $W = \TEC(p, r, r, r, t)$ is a balanced TEC with
    $H(W) = x$ and $E(W) = y$, then $x$ and $y$ uniquely determine $W$
    by
    \begin{align*}
        p & = 1 - x - \frac y2, \\
        r & = \frac y3, \\
        t & = x - \frac y2.  
    \end{align*}
    Using these, we can easily derive that
    \begin{align*}
        H(W^\para) & = x^2 - \frac{y^2}{12}, \\
        E(W^\para) & = 2xy - \frac{2y^2}3, \\
        H(W^\seri) & = 2x - x^2 + \frac{y^2}{12}, \\
        E(W^\seri) & = 2y - 2xy - \frac{2y^2}3.
    \end{align*}
    These expressions will be useful for this appendix.  Note that we
    use the balanced-ness condition here to express $H(W^\para)$,
    $E(W^\para)$, $H(W^\seri)$, and $E(W^\seri)$ using two parameters.

\subsection{Trapping region (Theorem \ref{thm:trap-Q})}\label{pf:trap-Q}

    Given $Q(W) \geq \alpha \coloneqq 2\sqrt7 - 4$, now we want to prove
    $Q(W^\para) \geq \alpha$ as $Q(W^\seri) \geq \alpha$ will follow by
    duality.

    \begin{IEEEproof}
        Consider a balanced TEC with entropy $H(W) = x$ and edge mass
        $E(W) = y$.  We first prove that $Q(W^\para) \geq \alpha$
        whenever $Q(W) = \alpha$, that is, whenever $y = \alpha x
        (1 - x)$.  To do so, we want $E(W^\para) \geq \alpha H(W^\para)
        (1 - H(W^\para))$;  so we want the following difference to be
        nonnegative:
        \begin{align*}
            \kern2em&\kern-2em
            E(W^\para) - \alpha H(W^\para) (1 - H(W^\para)) \\
            & = 2xy - \frac{2y^2}3
              - \alpha \Bigl( x^2 - \frac{y^2}{12} \Bigr)
                       \Bigl( 1 - x^2 + \frac{y^2}{12} \Bigr) \\
            & = \alpha x^2 (1 - x)^2 \cdot (X - Y)
        \end{align*}
        where $144X$ is
        \[ \alpha x^2 (1 - x) (24 + 24x + \alpha^2x^2 - \alpha^2x^3) \]
        and $12Y$ is
        \[ \alpha^2 + 8 \alpha - 12. \]
        $Y$ is zero because $2\sqrt7 - 4$ is a root.  $X$ is clearly
        nonnegative because $0 \leq x \leq 1$.  This confirms that
        $E(W^\para) \geq \alpha H(W^\para) (1 - H(W^\para))$ and that
        $Q(W^\para) \geq \alpha$ given that $y = \alpha x (1 - x)$.

        Now that we finished proving $Q(W^\para) \geq \alpha$ when $Q(W)
        = \alpha$, it remains to consider the case when $Q(W) > \alpha$,
        that is, when $y > \alpha x (1 - x)$.  Since the treatment
        for this case is lengthy and somewhat unrelated to the
        techniques used above, we put remaining of the proof in the next
        lemma.
    \end{IEEEproof}

    \begin{lemma}[already trapped]                \label{lem:already-QQ}
        If $Q(W_0) > \alpha$ for some balanced TEC $W_0$, there exists a
        balanced TEC $W_1$ such that $Q(W_0^\para) \geq Q(W_1^\para)
        \geq \alpha$, which witnesses $Q(W_0^\para) \geq \alpha$ and
        finishes the proof of \cref{thm:trap-Q}.
    \end{lemma}
    
    \begin{IEEEproof}
        Consider this map from $\RR^2$ to $\RR^2$
        \[
            \pi(x,y) \coloneqq \Bigl(\,
                x^2-\frac{y^2}{12}\,,\, 2xy-\frac{2y^2}3
            \,\Bigr)
        \]
        that encodes the evolution of $(H, E)$ under parallel
        combination.  We claim that, for any $(x_0, y_0)$-pair such that
        $y_0 > \alpha x_0 (1 - x_0)$, there exist a pair $(x_1, y_1)$
        such that
        \begin{itemize}
            \item $(x_0, y_0)$ lies above the parabola $y = \alpha x (1 -
                x)$,
            \item $(x_1, y_1)$ lies on the parabola $y = \alpha x (1 -
                x)$,
            \item $(x_0, y_0)$ and $(x_1, y_1)$ lie on the same
                hyperbola of the form $x^2 - y^2/12 = \text{const}$.
            \item $\pi(x_0, y_0)$ and $\pi(x_1, y_1)$ lie on the same
                vertical line, and
            \item $\pi(x_0, y_0)$ lies above $\pi(x_1, y_1)$.
        \end{itemize}    
        That is to say, there is a TEC $W_1$ that has the ``correct
        index'' $Q(W_1) = \alpha$ we know how to deal with; and we will
        use $W_1$ as a reference point to bound $Q(W_0^\para)$.
        
        To prove the claim, consider the ordinary differential equation
        (ODE):
        \begin{align*}
            f(0) & = x_0, & f'(t) = -1/f(t), \\
            g(0) & = y_0, & g'(t) = -12/g(t).
        \end{align*}
        The intuition behind this ODE is to imagine a particle starting
        from $(x_0, y_0)$ and traveling downward along a hyperbola $x^2
        - y^2/12 = \text{const}$ until $(f, g)$ reaches the parabola
        $y = \alpha x (1 - x)$.  The moment it reaches the parabola,
        the position this particle is at is the $(x_1, y_1)$ we want.
        To make this argument sound, there are several details we need
        to verify.

        (A) We need that the solution to the ODE exists til $g$ become
        $0$.  This is true because $E(W) \leq 2H(W)$ for any TEC $W$ and
        so we only need to consider the initial points $(x_0, y_0)$ that
        satisfy $y_0 \leq 2x_0$.  Now that $g$ is initialized as a value
        lower than $2f$'s and it decreases faster than $2f$ does, we
        conclude that $g \leq 2f$ and that $g$ will reach zero first,
        and before that happens the solution to the ODE is well-defined.

        (B) We need that the particle does travel along a hyperbola
        $x^2 - y^2/12 = \text{const}$.  This is true as the derivative
        \begin{align*}
            \Bigl( f^2 - \frac{g^2}{12} \Bigr)'
            & = 2ff' - \frac{gg'}{6} \\
            & = - \frac{2f}{f} + \frac{12g}{6g} \\
            & = 0
        \end{align*}
        does vanish.

        (C) We need that as the particle travels, $g/f(1-f)$ is
        decreasing. This is true because we can compute the derivative
        \begin{align*}
            \kern1em&\kern-1em
            \Bigl( \frac {g} {f (1 - f)} \Bigr)' \\
            & =  \frac {g' f (1 - f) - g f' (1 - f) + g f (1 - f')}
                       {f^2 (1 - f)^2} \\
            & =  \frac {- 12 f^2 (1 - f) + g^2 (1 - f) + g^2 f (f + 1)}
                       {g f^3 (1 - f)^2} \\
            & =  \frac {12f^3 - 12f^2 - 2fg^2 + g^2}
                       {g f^3 (1 - f)^2} \\
            & =  \frac {- 12 f^2 (1 - f) + g^2 (1 - 2f)}
                       {g f^3 (1 - f)^2}.
        \end{align*}
        The derivative is negative (or zero) because $12f^2 \geq g^2$
        and $1 - f \geq 1 - 2f$.

        (D) We need that the image of the particle is traveling
        downward.  This is true because the derivative of the vertical
        coordinate of $\pi(f, g)$:
        \begin{align*}
            \Bigl( 2fg - \frac{2g^2}{3} \Bigr)'
            & = 2f'g + 2fg' - \frac{4gg'}{3} \\
            & = - \frac{2g}{f} - \frac{24f}{g} + 16 \\
            & = \frac{- 2 (6f - g) (2f - g)} {fg} \\
            & \leq 0.
        \end{align*}
        is indeed negative or zero.

        Combining (A), (B), (C), and (D) proves the lemma.
    \end{IEEEproof}

    Remark: we apologize for the lengthy proof of \cref{lem:already-QQ}.
    As it turns out, moving the particle $(f, g)$ along a vertical line
    does not work.  Instead,  we have to move $\pi(f, g)$ along a
    vertical line.

\subsection{Attraction toward the trap (Theorem \ref{lem:attract-Q})}
                                                    \label{pf:attract-Q}

    We want to prove that $Q(W) \leq \alpha - \varepsilon$ implies
    $Q(W^\para) \geq Q(W) \bigl( 1 + (1 - H(W)) \delta \bigr)$.  And
    then $Q(W^\seri) \geq Q(W) \bigl( 1 + H(W) \delta \bigr)$ will
    follow by duality.

    \begin{IEEEproof}
        Let $W$ be a balanced TEC with entropy $H(W) = x$, edge mass
        $E(W) = y$, and Q-index $Q(W) = y / x(1-x) = b$.  Suppose
        $b < \alpha - \varepsilon$, then
        \begin{align*}
            \kern2em&\kern-2em
            E(W^\para) - b H(W^\para) (1 - H(W^\para)) \\
            & = b x^2 (1 - x)^2 \cdot (X - Y)
        \end{align*}
        where $144X = b^2 (1 - x) (24 + 24x + b^2x^2 - b^2x^3)$ and $12Y
        = b^2 + 8b - 12$.  We have seen that $144X$ is nonnegative.  For
        $Y$, we have $Y = (b - \alpha) (b - (-2\sqrt7 - 4)) < - 12
        \varepsilon (2\sqrt7 + 4) < - 9\varepsilon$.  Therefore,
        \begin{align*}
            Q(W^\para)
            & = \frac {E(W^\para)} {H(W^\para) (1 - H(W^\para))} \\
            & = b + \frac {bx^2 (1 - x)^2 (X - Y)}
                          {H(W^\para) (1 - H(W^\para))} \\
            & \geq b + \frac {b x^2 (1 - x)^2 (X - Y)}
                             {x^2 (1 - x^2)} \\
            & \geq b + \frac {b (1 - x) (X - Y)}
                             {1 + x} \\
            & \geq b + \frac {9 b (1 - x) \varepsilon} {12}.
        \end{align*}
        So $3\varepsilon/8$, our choice of $\delta$, is valid.
    \end{IEEEproof}

\subsection{Uniform attraction (Theorem \ref{thm:uniform-Q})}
                                                    \label{pf:uniform-Q}

    We want to prove that if $Q(W) \leq \alpha - \varepsilon$, there
    exists an integer $m > 0$ such that $Q(W_n) \geq Q(W)
    (1+\varepsilon/8)$ for all $n \geq m$.

    \begin{IEEEproof}
        In this proof, we call a descendant $W'$ of $W$ \emph{good} if
        $Q(W') \geq Q(W) (1+\varepsilon/8)$.  Being good is hereditary:
        if a balanced TEC is good, its descendants are all good because
        their $Q$'s, as long as $Q < \alpha$, are non-decreasing
        (\cref{lem:attract-Q}).  Now imagine the family tree consisting
        of the root $W$ and all the descendants that are not good.  The
        goal of this theorem is to show that this tree is finite.

        Fix a balanced TEC $W$, we know either $H(W) \geq 1/3$ or $H(W)
        \leq 2/3$.  For the former case,
        \begin{align*}
            Q(W^\seri)
            & \geq Q(W)\bigl(1+H(W)\delta\bigr) \\
            & \geq Q(W)\Bigl(1+\frac13\cdot\frac{3\varepsilon}8\Bigr) \\
            & \geq Q(W)\Bigl(1+\frac{\varepsilon}8\Bigr).
        \end{align*}
        For the latter case,
        \begin{align*}
            Q(W^\para)
            & \geq Q(W)\bigl(1+(1-H(W))\delta\bigr) \\
            & \geq Q(W)\Bigl(1+\frac13\cdot\frac{3\varepsilon}8\Bigr) \\
            & \geq Q(W)\Bigl(1+\frac{\varepsilon}8\Bigr).
        \end{align*}
        We infer that $H(W) \geq 1/3$ implies $W^\seri$ good and $H(W)
        \leq 2/3$ implies $W^\para$ good.  We see that the family tree
        of the bad descendants is uniparous---every node in this tree
        has at most one child.

        At this point, the only concern is whether there exists an
        infinite path of TECs $W, W_1, W_2,\allowbreak W_3,\allowbreak
        \ldots$ such that each is a child of the previous TEC, and none
        of them has entropy lying in $[1/3, 2/3]$.  To see why this
        cannot happen, suppose we begin with $x \coloneqq H(W) > 2/3$.
        We know $W^\seri$ is definitely good so $W_1$ must be $W^\para$.
        Given that $H(W^\para) = x^2 - y^2/12$ and $0 \leq y \leq 2 -
        2x$, we see
        \[
            H(W_1)
            = x^2 - \frac{y^2}{12}
            \geq x^2 - \frac{(1-x)^2}3
            \geq \frac{11}{27}
            \geq \frac13.
        \]
        This implies
        that the ``gap'' $[1/3, 2/3]$ is too large and that the path of
        TECs $W, W_1, W_2, \ldots$ cannot cross this gap---it must stay
        within $(2/3, 1)$ if $H(W)$ began there.

        According to the last few paragraphs, what will contradict the
        theorem is a path of TECs $W, W_1, \ldots$ such that each is the
        parallel-child of the previous and all of them have entropy $>
        2/3$.  But this cannot happen because the $H$ of the
        parallel-child is at most the square of the previous $H$, and
        squaring a number in $(2/3, 1)$ will eventually make it less
        than $2/3$.  By duality, there cannot be a path of TECs such
        that each is the serial-child of the previous and all of them
        have entropy $< 1/3$.  This finishes the proof.
    \end{IEEEproof}

\subsection{Attraction on the other side
            (Proposition \ref{pro:outer-Q})}          \label{pf:outer-Q}

    Given $Q(W) \leq 2$, we want to prove $Q(W^\para) \leq 2$ as
    $Q(W^\seri) \leq 2$ will follow by duality.

    \begin{IEEEproof}
        Consider a balanced TEC with entropy $H(W) = x$ and edge mass
        $E(W) = y = c x (1 - x)$ for some $c \leq 2$.  We prove
        $Q(W^\para) \leq 2$ by considering the positivity of the
        following quantity:
        \begin{align*}
            \kern0.5em&\kern-0.5em
            2 H(W^\para) (1 - H(W^\para)) - E(W^\para) \\
            & = \frac1{72} x^2 (1 - x) \cdot
                \bigl( 6 (2 - c) + c^2 x (1 - x) \bigr) \\
            &\kern0.5em
            \cdot \bigl(
                (6 + cx) (2 - c)
                + x (12 + 4c  + 2 c^2 x - c^2 x^2)
            \bigr).
        \end{align*}
        This is clearly nonnegative, which finishes the proof.
    \end{IEEEproof}

\subsection{Weak trap for the unbalanced
            (Theorem~\ref{thm:weak-Q})}                \label{pf:weak-Q}

    We will prove two claims.  (A) There exists a small number
    $\varepsilon > 0$ such that $Q(W) < \varepsilon$ implies $Q(W^\para)
    \geq Q(W)$.  (B) There exists a small number $\delta > 0$ such that
    $Q(W^\para) \leq \delta$ implies $Q(W) \leq \varepsilon$.  After
    proving (A) and (B), it is left to readers to check that
    $\min(\varepsilon, \delta)$ is a small number that meets the
    requirement of the theorem statement: $Q(W^\seri), Q(W^\para) \geq
    \min(Q(W), \min(\varepsilon, \delta))$

    \begin{IEEEproof}
        We first discuss (A).  If $Q(W) < \varepsilon$ for some very
        small $\varepsilon > 0$, then $E(W) < \varepsilon H(W) (1 -
        H(W)) < \varepsilon H(W) (1 - t) = \varepsilon (E(W)/2 + t) (1 -
        t)$.  This implies that
        \[
            E(W)
            < \frac {\varepsilon} {1 - \varepsilon/2} \cdot t (1 - t)
            < O(\varepsilon') t,
        \]
        where $\varepsilon' \coloneqq \varepsilon (1 - t)$.
        In other words, $t$ is very large compared to $q, r, s$.
        
        We now see
        \begin{align*}
            E(W^\para)
            & = qr + rs + sq + 2t E(W) \\
            & = 2t E(W) (1 \pm O(\varepsilon')).
        \end{align*}
        and
        \[ H(W) = \frac{E(W)}{2} + t = t (1 + O(\varepsilon'))\]
        and
        \begin{align*}
            H(W^\para)
            & = \frac{E(W^\para)}{2} + t^2 \\
            & = t^2 (1 \pm O(\varepsilon')) \\
            & = t H(W) (1 \pm O(\varepsilon'))
        \end{align*}
        and
        \begin{align*}
            1 - H(W^\para)
            & = 1 - t^2 (1 \pm O(\varepsilon')) \\
            & = (1 - t) (1 + t) (1 \pm O(\varepsilon')) \\
            & = (1 - H(W)) (1 + t) (1 \pm O(\varepsilon')).
        \end{align*}
        So,
        \begin{align*}
            Q(W^\para)
            & = \frac {E(W^\para)} {H(W^\para) (1 - H(W^\para))} \\
            & = \frac{2t E(W) (1 \pm O(\varepsilon'))}
                {t H(W) (1 - H(W)) (1 + t) (1 \pm O(\varepsilon'))} \\
            & = \frac{2 (1 \pm O(\varepsilon'))}{1 + t} \cdot Q(W),
        \end{align*}
        which is greater than $Q(W)$.  This finishes the proof of (A).

        We next discuss (B).  Suppose $Q(W^\para) \leq \delta$ for some
        very small $\delta > 0$.  Then by definition, $\delta \geq
        Q(W^\para) = E(W^\para) / H(W^\para) (1 - H(W^\para)) \geq
        E(W^\para) / H(W^\para) = E(W^\para) / (E(W^\para)/2 + t^2)$.
        This inequality suggests that $t^2$ is very large compared to
        $E(W^\para) = O(E(W))^2 + 2t E(W)$.  We then infer that $q, r,
        s$ is small compared to $t$ and that implies that $Q(W)$ is
        sufficiently small.  This finishes the proof of (B).
    \end{IEEEproof}

\subsection{Become extreme and then mediocre
            (Theorem~\ref{thm:reenter-H})}          \label{pf:reenter-H}

    We focus on how to select $0 < a < b$ to control the behavior of
    $\{H(W_n)\}_n$ after it gets below $a$ and climbs back to $b$.  The
    same argument applies, by symmetry, to the selection of $c < d < 1$.

    \begin{IEEEproof}
        Let's first discuss how $A(W_n)$ evolves.  First, by the second
        inequality of \cref{thm:monotone-A/E}, we know $A(W_n) / E(W_n)
        \leq 2E(W_n) \leq 4H(W_n) < 4a$ when $H(W_n)$ escapes $[a, d]$.
        By the first inequality of \cref{thm:monotone-A/E}, we also know
        that the $A/E$-ratio only decreases, so all $A(W_n) / E(W_n)$
        afterward are $< 4a$.  Hence, when $H(W_n)$ increases and
        crosses $\geq b$, we have $A(W_n) < 4a E(W_n) \leq 4a$.  We can
        make $A(W_n) / E(W_{n+1}) = A(W_n) / Q(W_{n+1}) H(W_{n+1}) (1 - 
        H(W_{n+1})) < 4a / \varepsilon b (1 - b)$ smaller than $\gamma$
        by choosing an $a$ smaller than $\gamma \varepsilon b (1 - b) /
        4$.  Here, this $\varepsilon$ is the $\varepsilon$ in the
        statement of \cref{thm:weak-Q}.

        Let's next discuss how $Q(W_n)$ evolves.
        For simplicity, let's first assume that $\{W_n\}_n$ undergoes
        $^{\para\para\dotsb\para\seri\seri\dotsb\seri}$
        to escape $[a, d]$ and then reenter $[b, c]$.
        
        When the channels are undergoing parallel combinations, we claim
        that $t$ will decay faster than $E$ will.  To see why, observe
        that $^\para$ evolves $q$ into $ts + sq + qt \geq (q + s) t$,
        which means that $^\para$ evolves $E$ into $\geq 2Et$.  But
        $^\para$ evolves $t$ into $t^2$ only.  In other words, the
        $E$-to-$t$ ratio doubles every parallel combination.  Hence
        $Q(W_n)$, which is $E / H (1 - H) \geq E/H = E/(E/2 + t) = 2 \pm
        O(t/E)$, will converge to $2$ quickly.  (Note that this is
        similar to \cref{lem:attract-Q}, where $Q(W^\para)$ increases by
        a larger amount when $H(W)$ is smaller.) To be more precise,
        either of the following happens when $H(W_n)$ is still $\geq b$:
        (A) The $E$-to-$t$ ratio is big enough such that $Q > 2 -
        \gamma$.  Since $^\para$ only increases $E/t$, $Q$ always stays
        $> 2 - \gamma$.  (B) The $E$-to-$t$ ratio is not big enough, in
        which case $E$ and $t$ satisfies $b \leq H = E/2 + t = (E/2t +
        1) t$ and hence $t$ is bounded from below.  Now $^\para$ evolves
        $t$ into $t$, $t^2$, $t^4$, etc and multiplies $E$ by $2t$, then
        by $2t^2$, then by $2t^4$, etc, which are all lower-bounded.  It
        is possible to choose a small $a/b$ so that, by the time
        $H(W_n)$ becomes $< a$, the $E/t$-ratio is doubled by a certain
        amount of times such that $Q > 2 - \gamma$.

        When the channels are undergoing serial combinations, we claim
        that $E$ and $t$ will evolve into $2E$ and $2t$, respectively,
        up to some multiplicative errors $1 \pm O(H)$.  To see why,
        observe that $^\seri$ evolves $q$ into $ps + sq + qp \geq p (q +
        s)$, where $p \geq 1 - 2H \approx 1$; similar estimates apply to
        $r$, $s$, and $t$.  (Note that this can be seen as the
        unbalanced version of the square-or-double behavior.)  From
        that, we see that the $E/t$-ratio remains somewhat unchanged,
        which implies that $Q$ remains somewhat unchanged and is $> 2 -
        O(\gamma)$.  (And then can make $Q > 2 - \gamma$ by replacing
        $\gamma$ with $\Omega(\gamma)$.)

        It remains to show that the order of parallel and serial
        combinations does not invalidate the claim that $Q$ will
        converge to $2 - \gamma$.  For the case (A) above, nothing will
        change.  For the case (B) above, if there is any $^\seri$
        inserted among $^{\para\para\dotsb\para}$, they only replace $t,
        t^2, t^4, \dotsc$ and $2t, 2t^2, 2t^4$ by greater numbers, so
        the argument for (B) still holds.  To see that the error term $1
        \pm O(H)$ will not alter the $E/t$-ratio by too much, we note
        that for any $^{\seri\seri\dotsb\seri}$ inserted between two
        $^\para$'s, $H(W_{n+1}) \approx 2H(W_n)$ and hence the error
        terms $1 \pm O(H) = \exp(\pm O(H))$ accumulate to $\exp\bigl(
        O(H(W_n) \pm 2H(W_n) \pm \dotsb) \bigr) = \exp(\pm O(b))$.
        Considering that a single $^\para$ will double $E/t$ and will
        shadow the effect of $\exp(\pm O(b))$, we conclude that
        the order of $^\para$'s and $^\seri$'s does not matter.

        To summarize the proof of \cref{thm:reenter-H}, we choose $b$
        (and $c \coloneqq 1 - b$) by making sure that $\exp(\pm O(b))$
        is close to $1$.  We than choose $a$ (and $d \coloneqq 1 - a$)
        by making sure that $A$ is small enough and $E/t$ doubles
        sufficiently many times so that $Q$ is close enough to $2$.
    \end{IEEEproof}

\section{Eigenvalue and Eigenvector (Lemma \ref{lem:eigen})}
                                                       \label{app:eigen}

    \Cref{lem:eigen} has $\psi(x) \coloneqq (x(1-x))^{0.697}
    \allowbreak (5 - \sqrt{x(1-x)})$ and $W$ balanced and $Q(W) \geq
    9/7$.  We want to verify that
    \[
        \frac {\psi(H(W^\seri)) + \psi(H(W^\para))}
              {2\psi(H(W))}
        < 0.818.
    \]

    Let $H(W) = x$ and $E(W) = y = 9x(1-x)/7$.  Then $H(W^\para) = x^2 -
    y^2/12 = (169x^2 + 54x^3 - 27x^4) / 196$, while $H(W^\seri) = 2x -
    x^2 + y^2/12 = (392x - 169x^2 - 54x^3 + 27x^4) / 196$.  The
    statement we want to prove boils down to showing
    \begin{equation}
        \frac {
            \psi \bigl( \frac {169x^2 + 54x^3 - 27x^4} {196} \bigr)
            + \psi \bigl( \frac {392x-169x^2-54x^3+27x^4} {196} \bigr)
        }
        {2\psi(x)} \label{for:quot}
    \end{equation}
    $< 0.818$ for $0 < x < 1$.  One can verify this numerically.

    If $y > 9x(1-x)/7$, then $H(W^\para)$ and $H(W^\seri)$ are more
    separated than when $y = 9x(1-x)/7$.  Hence $\psi(H(W^\seri)) +
    \psi(H(W^\para))$ is smaller than the numerator of
    formula~\eqref{for:quot} as $\psi$ is convex.  Hence the quotient is
    still smaller than $0.818$.

\subsection{Suboptimality of trapping region}

    The eigenvalue $0.818$ is not optimal in two aspects.  For one, the
    eigenfunction $\psi$ is not optimal.  A common practice is to run
    power iteration
    \begin{align*}
        \psi_0(x)     & \coloneqq (x(1-x))^{0.7}, \\
        \psi_{k+1}(x) & \coloneqq
        \frac {
            \psi_k \bigl(
                \frac {
                    \scalebox{0.7}[1]{$\scriptstyle
                        169x^2 + 54x^3 - 27x^4
                    $}
                } {196}
            \bigr)
            + \psi_k \bigl(
                \frac {
                    \scalebox{0.7}[1]{$\scriptstyle
                        392x - 169x^2 - 54x^3 + 27x^4
                    $}
                } {196} \bigr)
        } {2 \max \psi_k}
    \end{align*}
    until $\psi_k$ converges (note: use spline).  Then the limit of
    $\psi_k$ will induce a smaller eigenvalue.  See
    \cite[Section~III.C]{Sub4.7} for more details.

    For another, the trapping region we used is $Q(W) \geq 9/7$.  A
    better value is $\alpha = 2\sqrt7 - 4$ itself.  But even $Q(W) \geq
    2\sqrt7 - 4$ is not optimal.  A smaller trapping region is $y \geq x
    (1 - x) (1.66 - 0.38 x (1 - x))$.  But even that is not optimal.  So
    in the next appendix, we will use power iteration to find the
    optimal trapping region.

\section{Numerical Trapping Region}                  \label{app:iterate}

    In this appendix, we want to find the optimal (smallest) trapping
    region.  In this appendix, $W$ is a balanced TEC and $x = H(W)$ and
    $y = E(W)$.  Recall that $x$ and $y$ determine $W$ uniquely;  define
    \begin{align*}
        h_\para(x,y) \coloneqq H(W^\para) & = x^2-\frac{y^2}{12}, \\
        e_\para(x,y) \coloneqq E(W^\para) & = 2xy-\frac{2y^2}3, \\
        h_\seri(x,y) \coloneqq H(W^\seri) & = 2x-x^2+\frac{y^2}{12}, \\
        e_\seri(x,y) \coloneqq E(W^\seri) & = 2y-2xy-\frac{2y^2}3.  
    \end{align*}

    We call the lower boundary of a trapping region the \emph{inner
    bound} and the upper boundary of a trapping region the \emph{outer
    bound}.  For instance, $y = (2\sqrt7 - 4) x (1 - x)$ and $y = x (1 -
    x) (1.66 - 0.38 x (1 - x))$ are inner bounds; $y = 2 x (1 - x)$ and
    $y = x (1 - x) (2 - 2 x (1 - x) / 3)$ are outer bounds.
    Paraphrased, the mission of this appendix is to find the optimal
    inner and outer bounds.
    
\subsection{Numerical inner bound}

    Suppose $y = \varphi(x)$ is an inner bound, i.e., $\varphi$ is such
    that the children of a TEC above the curve $y = \varphi(x)$ will
    still be above the same curve.  Then the definition translates into:
    \begin{align*}
        e_\para(x,\varphi(x))
        & \geq \varphi \bigl( h_\para(x,\varphi(x)) \bigr), \\
        e_\seri(x,\varphi(x))
        & \geq \varphi \bigl( h_\seri(x,\varphi(x)) \bigr).  
    \end{align*}
    In other words,
    \begin{align*}
        \varphi(x) & \leq e_\para
        \bigl( h_\para^{-1}(x) , \varphi(h_\para^{-1}(x)) \bigr), \\
        \varphi(x) & \leq e_\seri
        \bigl( h_\seri^{-1}(x) , \varphi(h_\seri^{-1}(x)) \bigr).  
    \end{align*}
    where $h_\para^{-1}$ and $h_\seri^{-1}$ are inverse functions of $x
    \mapsto h_\para(x,\varphi(x))$ and $x \mapsto
    h_\seri(x,\varphi(x))$, respectively.
    
    Suppose there exists an optimal inner bound and it is of the form $y
    = \varphi(x)$, i.e., $\varphi$ is the greatest function such that
    the children of a TEC above the curve $y = \varphi(x)$ will still be
    above the same curve.  Then the optimality of $\varphi$ translates
    into
    \begin{align*}
        \varphi(x) = \min \Bigl(
        &e_\para\bigl(h_\para^{-1}(x),\varphi(h_\para^{-1}(x))\bigr),\\
        &e_\seri\bigl(h_\seri^{-1}(x),\varphi(h_\seri^{-1}(x))\bigr)
        \Bigr).
    \end{align*}

    We do not know a priori if the optimal inner bound exists.  But we
    can consider the following inductive definition
    \begin{align*}
        \varphi_0(x)   & \coloneqq 2 x (1-x), \\
        h_{\para k}(x) & \coloneqq h_\para(x, \varphi_k(x)), \\
        h_{\seri k}(x) & \coloneqq h_\seri(x, \varphi_k(x)), \\
        e_{\para k}(x) & \coloneqq e_\para \bigl( h_{\para k}^{-1}(x),
                               \varphi_k(h_{\para k}^{-1}(x)) \bigr), \\
        e_{\seri k}(x) & \coloneqq e_\seri \bigl( h_{\seri k}^{-1}(x),
                               \varphi_k(h_{\seri k}^{-1}(x)) \bigr), \\
        \varphi_{k+1}(x)&\coloneqq \min(e_{\para k}(x), e_{\seri k}(x)).
    \end{align*}
    These functions can be and were implemented by linear splines with
    $10^5$ nodes.  The advantage of linear splines is that the inverse
    function of a linear spline is still a linear spline. Per our
    computation, $\varphi_k$ converges as $k \to \infty$.  So we suspect
    that the limit of $\varphi_k$ is the optimal inner bound.
    
    A plot of the limit of $\varphi_k$ is in \cref{fig:trap}.

\subsection{Numerical outer bound}

    An argument similar to the previous sub-appendix applies to outer
    bound.  Suppose $y = \chi(x)$ is an outer bound, i.e., $\chi$ is
    such that the children of a TEC below the curve $y = \chi(x)$ will
    still be below the curve.  The definition translates into
    \begin{align*}
        e_\para(x, \chi(x))
        & \leq \chi \bigl( h_\para(x, \chi(x)) \bigr),\\
        e_\seri(x, \chi(x))
        & \leq \chi \bigl( h_\seri(x, \chi(x)) \bigr).  
    \end{align*}
    In other words,
    \begin{align*}
         \chi(x) & \geq e_\para
         \bigl( h_\para^{-1}(x), \chi(h_\para^{-1}(x)) \bigr), \\
         \chi(x) & \geq e_\seri
         \bigl( h_\seri^{-1}(x), \chi(h_\seri^{-1}(x)) \bigr).  
    \end{align*}
    
    Suppose $y = \chi(x)$ is the optimal outer bound, then the
    optimality implies
    \begin{align*}
        \chi(x) = \max \Bigl(
            & e_\para \bigl(
                h_\para^{-1}(x),
                \chi(h_\para^{-1}(x))
             \bigr), \\
            & e_\seri \bigl(
                h_\seri^{-1}(x),
                \chi(h_\seri^{-1}(x))
            \bigr)
        \Bigr).
    \end{align*}
    This means that we can setup an inductive definition almost
    identical to the one above except that the last line will be with
    $\max$:
    \begin{align*}
        \chi_0(x)      & \coloneqq 2 x (1-x), \\
        h_{\para k}(x) & \coloneqq h_\para(x, \chi_k(x)), \\
        h_{\seri k}(x) & \coloneqq h_\seri(x, \chi_k(x)), \\
        e_{\para k}(x) & \coloneqq e_\para \bigl( h_{\para k}^{-1}(x),
                                  \chi_k(h_{\para k}^{-1}(x)) \bigr), \\
        e_{\seri k}(x) & \coloneqq e_\seri \bigl( h_{\seri k}^{-1}(x),
                                  \chi_k(h_{\seri k}^{-1}(x)) \bigr), \\
        \chi_{k+1}(x) & \coloneqq \max(e_{\para k}(x), e_{\seri k}(x)).
    \end{align*}
    We did the computation and the end result of $\chi_k$ is plotted in
    \cref{fig:trap}.

\subsection{Improved estimate of scaling exponent}

    Using the numerical limit of $\varphi_k(x)$ in place of $(2\sqrt7 -
    4) x (1-x)$, we can strengthen \cref{lem:eigen,thm:main}.  Details
    omitted, our final number is $\mu < 3.328$.

    \begin{theorem}[main theorem with optimized constants]
        Consider a pair of BECs treated as a TEC, or consider any TEC
        where $pqrst > 0$.  The $2 \times 2$ matrix $\lowl$ over $\FF_4$
        induces a scaling exponent less than $3.328$.
    \end{theorem}

\pgfplotstableread{
    0.550000 -3.02900 0.828128 -5.33301 0.271872 -4.75399
    0.828128 -5.33301 0.978236 -12.2450 0.678020 -6.17317
    0.271872 -4.75399 0.483762 -6.12109 0.0599821 -9.92897
    0.978236 -12.2450 0.999678 -23.7651 0.956794 -12.3183
    0.678020 -6.17317 0.909752 -9.78724 0.446288 -7.45270
    0.483762 -6.12109 0.750275 -8.28478 0.217248 -8.49028
    0.0599821 -9.92897 0.117340 -10.1372 0.00262454 -18.5539
    0.999678 -23.7651 1.00000 -49.1091 0.999356 -23.7665
    0.956794 -12.3183 0.998650 -22.7406 0.914939 -12.4950
    0.909752 -9.78724 0.993741 -18.0088 0.825763 -10.1647
    0.446288 -7.45270 0.708568 -9.65633 0.184008 -10.4126
    0.750275 -8.28478 0.946406 -13.1827 0.554145 -9.38192
    0.217248 -8.49028 0.394603 -9.43105 0.0398931 -13.8461
    0.117340 -10.1372 0.223892 -10.6249 0.0107874 -17.4544
    0.00262454 -18.5539 0.00524459 -18.5651 4.49665e-6 -37.5289
    1.00000 -49.1091 1.00000 -97.4933 1.00000 -49.1091
    0.999356 -23.7665 1.00000 -45.7343 0.998712 -23.7688
    0.998650 -22.7406 0.999999 -42.6840 0.997301 -22.7455
    0.914939 -12.4950 0.994512 -20.4209 0.835365 -12.8183
    0.993741 -18.0088 0.999973 -33.4584 0.987509 -18.0313
    0.825763 -10.1647 0.975017 -15.8552 0.676508 -10.8424
    0.708568 -9.65633 0.925175 -13.7504 0.491962 -10.8591
    0.184008 -10.4126 0.339692 -11.1307 0.0283247 -15.9315
    0.946406 -13.1827 0.997826 -22.4079 0.894986 -13.3806
    0.554145 -9.38192 0.815396 -12.1246 0.292893 -11.4120
    0.394603 -9.43105 0.646883 -11.1673 0.142323 -12.5688
    0.0398931 -13.8461 0.0786034 -13.9927 0.00118287 -23.9403
    0.223892 -10.6249 0.405528 -11.5290 0.0422555 -15.5925
    0.0107874 -17.4544 0.0214936 -17.4941 0.0000811300 -31.4081
    0.00524459 -18.5651 0.0104705 -18.5841 0.0000186666 -34.5494
    4.49665e-6 -37.5289 8.99329e-6 -37.5289 1.30744e-11 -73.7537
    1.00000 -97.4933 1.00000 -196.566 1.00000 -97.4933
    1.00000 -49.1091 1.00000 -96.4212 1.00000 -49.1091
    1.00000 -45.7343 1.00000 -91.0451 0.999999 -45.7343
    0.998712 -23.7688 0.999999 -44.3422 0.997425 -23.7739
    0.999999 -42.6840 1.00000 -83.4720 0.999998 -42.6840
    0.997301 -22.7455 0.999995 -41.1029 0.994606 -22.7561
    0.994512 -20.4209 0.999979 -36.7249 0.989045 -20.4425
    0.835365 -12.8183 0.978115 -18.9314 0.692615 -13.4951
    0.999973 -33.4584 1.00000 -65.3512 0.999946 -33.4585
    0.987509 -18.0313 0.999891 -31.9289 0.975128 -18.0805
    0.975017 -15.8552 0.999549 -27.7053 0.950486 -15.9536
    0.676508 -10.8424 0.907058 -14.7779 0.445959 -12.2749
    0.925175 -13.7504 0.995642 -22.2517 0.854708 -14.0452
    0.491962 -10.8591 0.756003 -13.2669 0.227920 -13.3758
    0.339692 -11.1307 0.575708 -12.6388 0.103676 -14.8839
    0.0283247 -15.9315 0.0560603 -16.0423 0.000589220 -27.3701
    0.997826 -22.4079 0.999997 -41.4289 0.995656 -22.4165
    0.894986 -13.3806 0.991295 -20.8563 0.798676 -13.7993
    0.815396 -12.1246 0.971297 -17.8221 0.659495 -12.8802
    0.292893 -11.4120 0.509806 -12.6750 0.0759809 -15.6364
    0.646883 -11.1673 0.887172 -14.7850 0.406593 -12.7431
    0.142323 -12.5688 0.268028 -13.1418 0.0166188 -19.0844
    0.0786034 -13.9927 0.152448 -14.3037 0.00475880 -22.3593
    0.00118287 -23.9403 0.00236479 -23.9450 9.48099e-7 -44.7462
    0.405528 -11.5290 0.660833 -13.4085 0.150224 -14.7071
    0.0422555 -15.5925 0.0831897 -15.7580 0.00132126 -25.8303
    0.0214936 -17.4941 0.0426600 -17.5782 0.000327289 -29.7587
    0.0000811300 -31.4081 0.000162256 -31.4084 4.37657e-9 -59.9962
    0.0104705 -18.5841 0.0208655 -18.6254 0.0000755171 -33.0132
    0.0000186666 -34.5494 0.0000373329 -34.5495 2.29290e-10 -67.5025
    8.99329e-6 -37.5289 0.0000179865 -37.5289 5.35585e-11 -72.4892
    1.30744e-11 -73.7537 2.61488e-11 -73.7537 1.08727e-22 -147.928
    1.00000 -196.566 1.00000 -392.406 1.00000 -196.566
    1.00000 -97.4933 1.00000 -193.189 1.00000 -97.4933
    1.00000 -96.4212 1.00000 -190.117 1.00000 -96.4212
    1.00000 -49.1091 1.00000 -93.9201 0.999999 -49.1091
    1.00000 -91.0451 1.00000 -180.292 1.00000 -91.0451
    0.999999 -45.7343 1.00000 -88.2764 0.999999 -45.7343
    0.999999 -44.3422 1.00000 -84.8832 0.999998 -44.3422
    0.997425 -23.7739 0.999996 -42.0276 0.994855 -23.7836
    1.00000 -83.4720 1.00000 -164.147 1.00000 -83.4720
    0.999998 -42.6840 1.00000 -80.9865 0.999995 -42.6841
    0.999995 -41.1029 1.00000 -77.4032 0.999990 -41.1029
    0.994606 -22.7561 0.999980 -38.8967 0.989232 -22.7766
    0.999979 -36.7249 1.00000 -68.8806 0.999959 -36.7250
    0.989045 -20.4425 0.999917 -34.5077 0.978173 -20.4839
    0.978115 -18.9314 0.999659 -30.9645 0.956572 -19.0143
    0.692615 -13.4951 0.917303 -17.5310 0.467927 -14.8185
    1.00000 -65.3512 1.00000 -128.143 1.00000 -65.3512
    0.999946 -33.4585 1.00000 -62.8120 0.999892 -33.4588
    0.999891 -31.9289 1.00000 -59.2741 0.999782 -31.9293
    0.975128 -18.0805 0.999560 -29.7371 0.950695 -18.1748
    0.999549 -27.7053 1.00000 -50.9365 0.999097 -27.7070
    0.950486 -15.9536 0.998181 -25.5498 0.902790 -16.1416
    0.907058 -14.7779 0.993206 -22.4685 0.820910 -15.1347
    0.445959 -12.2749 0.707251 -14.3415 0.184667 -15.0628
    0.995642 -22.2517 0.999987 -38.9411 0.991297 -22.2680
    0.854708 -14.0452 0.982759 -20.3822 0.726657 -14.6155
    0.756003 -13.2669 0.948216 -17.9753 0.563790 -14.2649
    0.227920 -13.3758 0.410986 -14.3006 0.0448544 -18.2979
    0.575708 -12.6388 0.833380 -15.5832 0.318036 -14.5712
    0.103676 -14.8839 0.198762 -15.2828 0.00859065 -22.2363
    0.0560603 -16.0423 0.109751 -16.2554 0.00236933 -25.2641
    0.000589220 -27.3701 0.00117821 -27.3723 2.34769e-7 -49.8466
    0.999997 -41.4289 1.00000 -78.9005 0.999994 -41.4289
    0.995656 -22.4165 0.999987 -39.1368 0.991325 -22.4328
    0.991295 -20.8563 0.999947 -35.5555 0.982642 -20.8891
    0.798676 -13.7993 0.965999 -19.1478 0.631354 -14.6140
    0.971297 -17.8221 0.999391 -29.0049 0.943203 -17.9298
    0.659495 -12.8802 0.895511 -16.5383 0.423479 -14.3520
    0.509806 -12.6750 0.773465 -15.1410 0.246146 -15.0110
    0.0759809 -15.6364 0.147438 -15.9254 0.00452369 -23.9209
    0.887172 -14.7850 0.989720 -21.8752 0.784624 -15.2202
    0.406593 -12.7431 0.660733 -14.5706 0.152454 -15.8223
    0.268028 -13.1418 0.473228 -14.2559 0.0628276 -17.5672
    0.0166188 -19.0844 0.0330383 -19.1467 0.000199325 -31.8754
    0.152448 -14.3037 0.286021 -14.9078 0.0188751 -20.5198
    0.00475880 -22.3593 0.00950195 -22.3772 0.0000156457 -38.8159
    0.00236479 -23.9450 0.00472578 -23.9539 3.81043e-6 -42.4249
    9.48099e-7 -44.7462 1.89620e-6 -44.7462 5.96873e-13 -85.7405
    0.660833 -13.4085 0.896826 -17.0963 0.424839 -14.8813
    0.150224 -14.7071 0.281959 -15.2997 0.0184884 -20.9526
    0.0831897 -15.7580 0.161071 -16.0792 0.00530873 -23.8270
    0.00132126 -25.8303 0.00264133 -25.8353 1.18426e-6 -46.0034
    0.0426600 -17.5782 0.0840007 -17.7415 0.00131931 -27.6691
    0.000327289 -29.7587 0.000654506 -29.7599 7.17625e-8 -53.9726
    0.000162256 -31.4084 0.000324494 -31.4091 1.75950e-8 -57.6361
    4.37657e-9 -59.9962 8.75314e-9 -59.9962 1.26974e-17 -116.492
    0.0208655 -18.6254 0.0414268 -18.7046 0.000304179 -30.8155
    0.0000755171 -33.0132 0.000151030 -33.0135 3.79975e-9 -61.4246
    0.0000373329 -34.5495 0.0000746649 -34.5496 9.27115e-10 -64.9685
    2.29290e-10 -67.5025 4.58580e-10 -67.5025 3.46508e-20 -132.424
    0.0000179865 -37.5289 0.0000359728 -37.5290 2.15496e-10 -70.0981
    5.35585e-11 -72.4892 1.07117e-10 -72.4892 1.89978e-21 -141.674
    2.61488e-11 -73.7537 5.22976e-11 -73.7537 4.50608e-22 -144.939
    1.08727e-22 -147.928 2.17455e-22 -147.928 7.63566e-45 -294.553
}\tableAtree
\pgfplotstabletranspose\tableAtransp\tableAtree
\pgfplotstablegetcolsof\tableAtransp

\begin{figure}\centering
    \iflabor
    \pgfplotstabletranspose\tableAtransp\tableAtree
    \pgfplotstablegetcolsof\tableAtransp
    \pgfmathtruncatemacro\ncol{\pgfplotsretval - 1}
    \def\readsegment#1#2#3#4#5#6#7#8{
        \readsegmentaux{#2}{#4}{#6}{#8}
    }
    \def\readsegmentaux#1#2#3#4#5#6#7#8{
        \draw (#1, #2) -- (#3, #4) (#1, #2) -- (#6, #8);
    }
    \begin{tikzpicture}
        \begin{axis}[mark size=.4pt]
            \addplot[opacity=0] coordinates {(0, -2) (1, -20)};
            \pgfplotsinvokeforeach{1,...,\ncol}{
                \pgfplotstablegetcolumn{[index]#1}\of\tableAtransp\to\CC
                \expandafter\readsegment\CC
            }
        \end{axis}
    \end{tikzpicture}
    \fi
    \caption{
        Horizontal axis: $H(W_n)$;  vertical axis: $\log_2(A(W_n))$.
        Tree: a pair of $\BEC(0.55)$ and its descendants up to the
        $7$th-generation.
    }                                                   \label{fig:tree}
\end{figure}
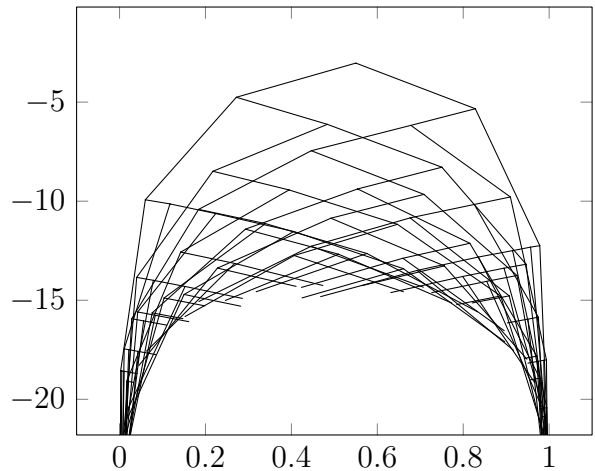

\section{Immediate Balance}                        \label{app:immediate}

    In \cref{app:balance}, we spend several paragraphs to explain how
    $A(W_n)$ will decay to $0$ as $n$ goes to infinity.  In this
    appendix, we present a coding technique that will balance any TEC at
    $n = 0$.

    Suppose $W = \TEC(p, q, r, s, t)$ is an unbalanced TEC that we
    want to send message over.  We build a new channel $\tilde W$ in the
    following way:  Upon receiving the input $(x_1, x_2) \in \FF_2^2$,
    pick a random matrix $M \in \{ \lool, \olll, \lllo \}$ independently
    and uniformly; input $xM$ into $W$; and then output $M$ and the
    output of $W$.

    Since $\tilde W$ is a derivation of $W$, it is a channel that is at
    most as good as $W$.  Meanwhile, we permute the subspace erasure
    probabilities $q$, $r$, and $s$ by a random power of the permutation
    $q \to r \to s \to q$.  Therefore, the probability that the receiver
    learns $x_1$ but not $x_2$ is the average $(q + r + s)/3$.  The same
    logic applies to other erasure patterns.  This implies that $\tilde
    W$ is at least as good as
    \[
        \bar W \coloneqq \TEC \Bigl(
            p,
            \frac{q + r + s}{3},
            \frac{q + r + s}{3},
            \frac{q + r + s}{3},
            t
        \Bigr).
    \]
    One then computes and sees that $H(\tilde W) = H(\bar W) = H(W)$.
    Thus, we did not lose any capacity by transmitting information over
    $\tilde W$, which we know is equivalent to a balanced TEC.

\pgfplotstableread{
    x1     y1     x2     y2     x3     y3     x4     y4     x5     y5
    .00028 .00057 .00059 .00119 .00120 .00240 .00243 .00483 .00490 .00967
    .00990 .01926 .01989 .03749 .29669 .37769 .00259 .00514 .00522 .01032
    .01052 .02054 .02116 .04035 .00031 .00063 .04215 .07660 .08060 .13463
    .52765 .43055 .02099 .04014 .00015 .00031 .00031 .00062 .04192 .07707
    .00063 .00125 .00127 .00251 .08233 .14030 .00487 .00936 .15397 .23008
    .25833 .32305 .78714 .28456 .00098 .00194 .00197 .00389 .00397 .00773
    .14096 .21911 .00749 .01433 .01488 .02758 .25634 .32664 .04562 .07766
    .41430 .39645 .56761 .40032 .94942 .08526 .00039 .00078 .04616 .08014
    .08599 .13629 .53507 .40445 .19417 .25835 .71708 .32958 .82627 .24387
    .99355 .01215 .39369 .39295 .88088 .18077 .93479 .11048 .99919 .00158
    .96637 .06115 .99980 .00040 .00012 .00024 .00024 .00049 .03737 .06966
    .00012 .00024 .00025 .00050 .00050 .00100 .00101 .00201 .07424 .13024
    .00101 .00202 .00204 .00403 .00412 .00803 .14380 .22454 .00783 .01502
    .01558 .02894 .26228 .33411 .04796 .08174 .42374 .40221 .57819 .40179
    .95257 .08092 .00159 .00314 .00320 .00628 .00642 .01245 .01285 .02425
    .24182 .32361 .01213 .02306 .02407 .04419 .04653 .08027 .41974 .40724
    .00107 .00209 .07445 .12248 .13375 .19545 .63069 .37676 .27152 .32175
    .79623 .26722 .88098 .18173 .99712 .00551 .00054 .00107 .00109 .00214
    .07569 .12655 .00407 .00776 .14024 .20732 .23535 .29464 .76263 .29628
    .02387 .04218 .31273 .34896 .45618 .39446 .90950 .14000 .63491 .38028
    .96617 .05899 .98257 .03204 .11636 .17473 .59945 .38802 .73590 .32353
    .98384 .02943 .84852 .22608 .99528 .00906 .99764 .00459 .91988 .13578
    .99879 .00239 .99940 .00119 .99970 .00059 .00022 .00045 .03578 .06539
    .00045 .00090 .00092 .00181 .06999 .11945 .00349 .00671 .13102 .19894
    .22284 .28796 .75078 .30970 .00429 .00825 .00855 .01607 .19866 .27233
    .02743 .04812 .33249 .36017 .47878 .39809 .91895 .12769 .10274 .15623
    .57119 .39107 .71125 .33631 .98013 .03558 .83140 .24240 .99401 .01137
    .99700 .00579 .00039 .00077 .04592 .07937 .08540 .13477 .53301 .40185
    .19212 .25482 .71404 .32940 .82363 .24499 .99331 .01255 .38908 .38841
    .87801 .18318 .93291 .11268 .99914 .00168 .96528 .06270 .99978 .00043
    .61295 .40397 .96197 .06730 .98054 .03635 .99025 .01876 .99513 .00950
    .01507 .02762 .25639 .31979 .39178 .38750 .87925 .18097 .57957 .40540
    .95316 .08049 .97573 .04441 .75187 .32781 .98636 .02564 .99317 .01320
    .99660 .00666 .86717 .21272 .99654 .00677 .99828 .00339 .99915 .00169
    .99957 .00084 .93207 .12072 .99916 .00167 .99958 .00083 .99979 .00041
    .00939 .01827 .01884 .03590 .00012 .00025 .00025 .00050 .03752 .06865
    .00100 .00198 .07300 .12387 .13309 .20081 .63301 .39005 .00020 .00041
    .00041 .00083 .00084 .00166 .06725 .11673 .00165 .00323 .00331 .00641
    .12856 .19968 .01170 .02167 .22874 .29763 .35700 .37364 .85913 .20459
    .01306 .02436 .02558 .04562 .32402 .36339 .07169 .11486 .49660 .39931
    .64518 .37043 .96811 .05534 .20338 .26725 .72814 .32249 .83429 .23579
    .99418 .01100 .90944 .14805 .99841 .00310 .99921 .00156 .00610 .01176
    .01217 .02297 .02398 .04341 .31654 .36665 .00030 .00059 .04028 .07039
    .07549 .12121 .50791 .40401 .17433 .23727 .69107 .34318 .80667 .26142
    .99183 .01522 .00545 .01029 .16041 .22821 .26366 .31396 .78885 .27202
    .44141 .39703 .90345 .14931 .94761 .08934 .99948 .00101 .64221 .38588
    .96810 .05661 .98368 .03043 .99180 .01570 .05654 .09426 .45266 .40366
    .60558 .39123 .95944 .06978 .75860 .31510 .98699 .02426 .99346 .01254
    .86810 .20791 .99655 .00672 .99828 .00338 .99915 .00169 .93187 .11966
    .99914 .00169 .99957 .00084 .99979 .00042 .00121 .00238 .00244 .00474
    .11114 .17668 .00875 .01636 .20026 .27130 .31944 .35361 .83408 .23136
    .04439 .07529 .40882 .39123 .56093 .39757 .94726 .08795 .72337 .33784
    .98223 .03248 .99101 .01703 .17548 .24401 .69478 .34982 .81129 .26342
    .99236 .01441 .89650 .16841 .99792 .00407 .99896 .00204 .94656 .09541
    .99948 .00103 .99974 .00051 .99987 .00025 .41559 .41191 .89360 .16860
    .94295 .10008 .99940 .00118 .97100 .05412 .99985 .00029 .98551 .02793
    .99279 .01411 .65529 .40071 .97164 .05218 .98575 .02734 .99292 .01386
    .99649 .00694 .99825 .00346 .99914 .00171 .00091 .00179 .06927 .11668
    .12600 .18937 .61897 .38883 .26318 .32146 .79022 .27733 .87787 .18861
    .99699 .00579 .47994 .42073 .92137 .12947 .95853 .07431 .99969 .00062
    .97901 .03954 .69119 .37906 .97795 .04103 .98895 .02131 .99451 .01077
    .99727 .00539 .83484 .25711 .99456 .01062 .99731 .00532 .99867 .00265
    .99934 .00132 .99967 .00065 .91610 .14789 .99871 .00256 .99936 .00127
    .99968 .00063 .99984 .00031 .95935 .07635 .99971 .00057 .99986 .00028
}\tableHE
\pgfplotstableread{
    x      1-x    out_math  out_iter  in_iter   in_math
    0.00   1.00   0.00000   0.00000   0.00000   0.00000   
    0.01   0.99   0.01980   0.01973   0.01835   0.012786   
    0.02   0.98   0.03920   0.03894   0.03547   0.025313   
    0.03   0.97   0.05820   0.05763   0.05180   0.037583   
    0.04   0.96   0.07680   0.07580   0.06749   0.049594   
    0.05   0.95   0.09500   0.09347   0.08262   0.061346   
    0.06   0.94   0.11280   0.11065   0.09716   0.072841   
    0.07   0.93   0.13020   0.12732   0.11121   0.084077   
    0.08   0.92   0.14720   0.14351   0.12481   0.095055   
    0.09   0.91   0.16380   0.15921   0.13797   0.10577   
    0.10   0.90   0.18000   0.17444   0.15073   0.11624   
    0.11   0.89   0.19580   0.18920   0.16309   0.12644   
    0.12   0.88   0.21120   0.20350   0.17507   0.13638   
    0.13   0.87   0.22620   0.21733   0.18668   0.14607   
    0.14   0.86   0.24080   0.23071   0.19793   0.15550   
    0.15   0.85   0.25500   0.24364   0.20883   0.16467   
    0.16   0.84   0.26880   0.25612   0.21939   0.17358   
    0.17   0.83   0.28220   0.26816   0.22962   0.18223   
    0.18   0.82   0.29520   0.27977   0.23952   0.19063   
    0.19   0.81   0.30780   0.29095   0.24909   0.19876   
    0.20   0.80   0.32000   0.30170   0.25835   0.20664   
    0.21   0.79   0.33180   0.31202   0.26729   0.21426   
    0.22   0.78   0.34320   0.32193   0.27593   0.22162   
    0.23   0.77   0.35420   0.33143   0.28426   0.22873   
    0.24   0.76   0.36480   0.34052   0.29222   0.23557   
    0.25   0.75   0.37500   0.34920   0.29974   0.24216   
    0.26   0.74   0.38480   0.35747   0.30695   0.24849   
    0.27   0.73   0.39420   0.36535   0.31385   0.25456   
    0.28   0.72   0.40320   0.37284   0.32046   0.26037   
    0.29   0.71   0.41180   0.37993   0.32677   0.26592   
    0.30   0.70   0.42000   0.38664   0.33278   0.27122   
    0.31   0.69   0.42780   0.39296   0.33850   0.27625   
    0.32   0.68   0.43520   0.39890   0.34392   0.28103   
    0.33   0.67   0.44220   0.40446   0.34906   0.28555   
    0.34   0.66   0.44880   0.40965   0.35391   0.28981   
    0.35   0.65   0.45500   0.41446   0.35847   0.29382   
    0.36   0.64   0.46080   0.41891   0.36274   0.29756   
    0.37   0.63   0.46620   0.42299   0.36673   0.30105   
    0.38   0.62   0.47120   0.42670   0.37044   0.30428   
    0.39   0.61   0.47580   0.43006   0.37386   0.30725   
    0.40   0.60   0.48000   0.43306   0.37700   0.30996   
    0.41   0.59   0.48380   0.43570   0.37986   0.31241   
    0.42   0.58   0.48720   0.43799   0.38244   0.31461   
    0.43   0.57   0.49020   0.43993   0.38473   0.31655   
    0.44   0.56   0.49280   0.44152   0.38675   0.31823   
    0.45   0.55   0.49500   0.44276   0.38848   0.31965   
    0.46   0.54   0.49680   0.44366   0.38994   0.32081   
    0.47   0.53   0.49820   0.44422   0.39111   0.32171   
    0.48   0.52   0.49920   0.44444   0.39200   0.32236   
    0.49   0.51   0.49980   0.44432   0.39262   0.32275   
    0.50   0.50   0.50000   0.44387   0.39295   0.32288   
}\tablephichi
\begin{figure}\centering
    \iflabor
    \begin{tikzpicture}
        \begin{axis}[mark size=.4pt]
            \addplot[only marks]table[x=x1,y=y1]{\tableHE};
            \addplot[only marks]table[x=x2,y=y2]{\tableHE};
            \addplot[only marks]table[x=x3,y=y3]{\tableHE};
            \addplot[only marks]table[x=x4,y=y4]{\tableHE};
            \addplot[only marks]table[x=x5,y=y5]{\tableHE};
            \addplot[PMS3015]table[x=x,  y=out_math]{\tablephichi};
            \addplot[PMS3015]table[x=1-x,y=out_math]{\tablephichi};
            \addplot[PMS1245]table[x=x,  y=out_iter]{\tablephichi};
            \addplot[PMS1245]table[x=1-x,y=out_iter]{\tablephichi};
            \addplot[PMS1245]table[x=x,  y=in_iter ]{\tablephichi};
            \addplot[PMS1245]table[x=1-x,y=in_iter ]{\tablephichi};
            \addplot[PMS3015]table[x=x,  y=in_math ]{\tablephichi};
            \addplot[PMS3015]table[x=1-x,y=in_math ]{\tablephichi};
        \end{axis}
    \end{tikzpicture}
    \fi
    \caption{
        Horizontal axis: $x = H(W_n)$;  vertical axis: $y = E(W_n)$.
        Curves from top to bottom: the provable outer bound $2 x (1-x)$,
        the numerical outer bound $\chi_{293}$, the numerical inner
        bound $\varphi_{120}$, and the provable inner bound $(2\sqrt7 -
        4) x (1 - x)$.  Dots: the $10$th-generation descendants of
        a pair of $\BEC(0.55)$.
    }                                                   \label{fig:trap}
\end{figure}
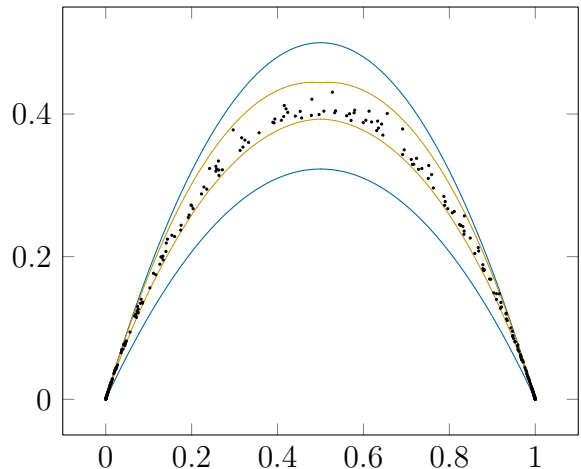

\section{Simulations}                               \label{app:simulate}

    In \cref{fig:tree}, we take a pair of $\BEC(0.55)$ and treat them as
    a TEC $W$.  We then compute $A(W_n)$ for all possible $W_n$ for $n
    \leq 7$.  In this figure, we see that $A$ decreases by a lot if $H$
    is going toward the closer end of $0$ or $1$, which is correctly
    predicted by \cref{thm:fast-A}.

    Note that, on paper, we can only prove that the worst-case decay of
    $A$ is about $O(1/n)$ (\cref{cor:ultimate-A}).  But in practice, we
    see that even the slowest-decaying branch decays exponentially fast.
    Per our simulation, the \emph{sum} of all $A(W_n)$ for $n$ fixed
    decays as fast as $C \cdot 1.95^{-n}$.

    In \cref{fig:trap}, we compute $H(W_n)$ and $E(W_n)$ for all
    possible $W_n$ for $n = 10$, and compare it with trapping regions.
    As can be seen in the figure, not only do points fall inside the
    provable trapping region (blue), they all fall inside the numerical
    trapping region (gold).

    Note that, on paper, we have to wait for $W_n$ to become
    almost-balanced and only when TECs are balanced enough can we infer
    that the Q-index will converge to the trapping region.  But in
    practice, we see that TECs become balanced and edge-heavy at the
    same time.

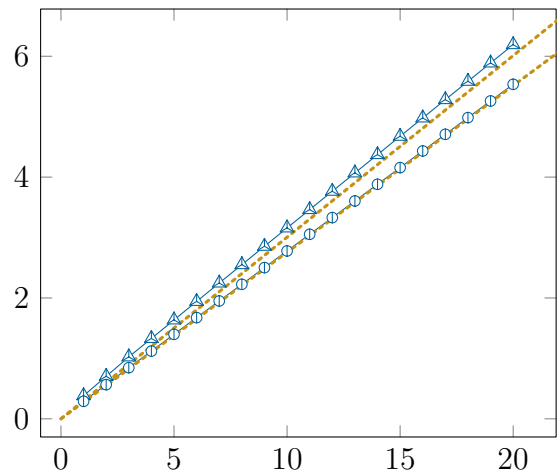
\begin{figure}\centering
    \iflabor
    \pgfdeclareplotmark{tetra}{%
        \pgfplotmarksize=1.5\pgfplotmarksize
        \pgfpathmoveto{\pgfqpointpolar{90}{\pgfplotmarksize}}%
        \pgfpathlineto{\pgfqpointpolar{-30}{\pgfplotmarksize}}%
        \pgfpathlineto{\pgfqpointpolar{210}{\pgfplotmarksize}}%
        \pgfpathlineto{\pgfqpointpolar{90}{\pgfplotmarksize}}%
        \pgfsetfillcolor{white}%
        \pgfusepathqfillstroke
        \pgfpathmoveto{\pgfqpointpolar{90}{\pgfplotmarksize}}%
        \pgfpathlineto{\pgfpointorigin}%
        \pgfpathmoveto{\pgfqpointpolar{-30}{\pgfplotmarksize}}%
        \pgfpathlineto{\pgfpointorigin}%
        \pgfpathmoveto{\pgfqpointpolar{210}{\pgfplotmarksize}}%
        \pgfpathlineto{\pgfpointorigin}%
        \pgfusepathqstroke
    }
    \pgfdeclareplotmark{binary}{%
        \pgfpathmoveto{\pgfqpoint{0pt}{-\pgfplotmarksize}}%
        \pgfpathlineto{\pgfqpoint{0pt}{\pgfplotmarksize}}%
        \pgfpathcircle{\pgfpointorigin}{\pgfplotmarksize}%
        \pgfsetfillcolor{white}%
        \pgfusepathqfillstroke
    }
    \pgfplotstableread{
        n    tec      bec
        1    0.3825   0.2898
        2    0.7031   0.5655
        3    1.0244   0.8465
        4    1.3296   1.1205
        5    1.6362   1.3984
        6    1.9426   1.6745
        7    2.2470   1.9503
        8    2.5509   2.2262
        9    2.8550   2.5022
        10   3.1585   2.7780
        11   3.4619   3.0538
        12   3.7654   3.3296
        13   4.0687   3.6054
        14   4.3721   3.8811
        15   4.6753   4.1569
        16   4.9784   4.4326
        17   5.2816   4.7084
        18   5.5848   4.9841
        19   5.8880   5.2598
        20   6.1912   5.5356
    }\tableslope
    \begin{tikzpicture}
        \begin{axis}
            \addplot[PMS3015,mark=tetra]table[x=n,y=tec]{\tableslope};
            \addplot[PMS3015,mark=binary]table[x=n,y=bec]{\tableslope};
            \draw[dotted,line width=1.2,PMS1245](0,0)--(50,50/3.328);
            \draw[dotted,line width=1.2,PMS1245](0,0)--(50,50/3.627);
        \end{axis}
    \end{tikzpicture}
    \fi
    \caption{
        Horizontal axis: generation (that is, $n$);  vertical axis:
        $-\log_2$ of the expectation of $\psi(H(W_n)) / \psi(H(W_0))$,
        where $\psi(x) \coloneqq (x(1-x)) ^ {0.7}$.  Triangle marks:
        $\BEC(0.55)$ polarized by $\lowl$; circle marks: $\BEC(0.55)$
        polarized by $\loll$.  Dotted lines: the lines of slopes
        $1/3.328$ (above) and $1/3.627$ (below).  This reaffirms that
        $\mu \approx 3.627$ is accurate for $\loll$ and $\mu < 3.328$ is
        an overestimate (but probably very close to the true value) for
        $\lowl$.
     }                                                 \label{fig:slope}
\end{figure}

    In \cref{fig:slope}, we compute the expectation of $\psi(H(W_n))$
    for $W_n$ defined by $\loll$ and that defined by $\lowl$, and
    compare them against lines with slopes $1/3.328$ and $1/3.627$.  The
    result shows a clear advantage of $\lowl$, which assumes
    a higher slope, over $\loll$.

    Note that, on paper, we have to wait for $W_n$ to become
    almost-balanced and only then the Q-index will converge to the
    trapping region, and then we have to wait for $W_n$ to become
    edge-heavy and only then we will see the improved scaling exponent.
    But in practice, all the above happen at the same time.

\bibliographystyle{IEEEtran}
\bibliography{TetraErase-14.bib}

\end{document}